\documentclass[aps,prx,twocolumn,superscriptaddress]{revtex4-2}
\usepackage[utf8]{inputenc}
\usepackage{graphicx}
\usepackage{epstopdf}
\usepackage{bm}
\usepackage{amsmath}
\usepackage{verbatim}
\usepackage{amssymb}
\usepackage{listings}
\usepackage{xcolor}
\usepackage{placeins}
\definecolor{darkblue}{rgb}{0,0,0.6}
\usepackage[colorlinks,linkcolor=darkblue,citecolor=darkblue,urlcolor=darkblue]{hyperref}
\usepackage{mathtools}

\newcommand{\is}{inherent state }
\newcommand{\Is}{Inherent state }
\newcommand{\iss}{inherent states }

\begin{document}
	
	\title{Numerical investigation of the equilibrium Kauzmann transition in a two-dimensional atomistic glass}
	
	\author{Gerhard Jung}
	
	\affiliation{Laboratoire Interdisciplinaire de Physique (LIPhy), Université Grenoble Alpes, 38402 Saint-Martin-d'Hères, France}
	
	\author{Misaki Ozawa}
	
	\affiliation{Laboratoire Interdisciplinaire de Physique (LIPhy), Université Grenoble Alpes, 38402 Saint-Martin-d'Hères, France}
	
	\author{Giulio Biroli}
	
	\affiliation{Laboratoire de Physique de l’Ecole Normale Supérieure, ENS, Université PSL, CNRS, Sorbonne Université, Université de Paris, 75005 Paris, France}
	
	\author{Ludovic Berthier}
	
	\affiliation{Gulliver, CNRS UMR 7083, ESPCI Paris, PSL Research University, 75005 Paris, France}
	
	\date{\today}
	
	\begin{abstract}
		Dense liquids gradually transform into non-equilibrium amorphous solids as they pass through the experimental glass transition.
		Experimentally, ergodicity is lost because measurements are conducted within a finite time window. More than seventy years ago, Kauzmann posed a fundamental question: If experiments could run indefinitely, would there exist a critical temperature at which an ergodicity-breaking phase transition occurs? Random first-order transitions represent the modern theoretical framework for this idea. However, theoretical calculations in finite dimensions are challenging, whereas experimental and numerical limitations on accessible timescales hinder direct observation of the putative Kauzmann transition. Here, we overcome this longstanding barrier by developing a computational strategy that properly combines three Monte Carlo methods to access the desired equilibrium thermodynamic properties of a two-dimensional atomistic glass-former down to zero temperature across a range of system sizes up to 77 particles. This enables us to directly measure thermodynamic and structural observables that provide unambiguous evidence that the system undergoes a finite-size version of the Kauzmann transition at a temperature that is much lower than other energy scales and decreases rapidly with system size, thus suggesting the existence of a zero-temperature Kauzmann transition for two-dimensional glasses. The transition is towards an ideal glass state characterized by a complex energy landscape and a hierarchical organization of low-lying states. 
	\end{abstract}
	
	\maketitle
	
	\section*{Introduction} 
	
	Conventional glasses are obtained when liquids fall out of equilibrium below the experimental glass transition temperature $T_g$~\cite{ediger1996supercooled}, see Fig.~\ref{fig:kauzmann}a. Below $T_g$, relaxational dynamics becomes slower than the observation time, and the liquid appears frozen in a non-ergodic amorphous state. In a landmark paper~\cite{kauzmann1948nature}, Kauzmann famously articulated fundamental questions regarding the fate of the liquid state if thermodynamic equilibrium and ergodicity could be maintained far below $T_g$. While Kauzmann debated the relative stability of liquid and crystalline states~\cite{tanaka2003possible}, his data collection also hinted at a possible thermodynamic phase transition at the (now called~\cite{stillinger1988supercooled,speedy2003kauzmann}) Kauzmann temperature, $T_K < T_g$, between a liquid and an ideal glass state, see Fig.~\ref{fig:kauzmann}a.
	
	The hypothesis of an experimentally inaccessible Kauzmann transition associated with an entropy crisis has been a central theme in glass research over the last decades~\cite{gibbs1958nature,stillinger1988supercooled,ediger1996supercooled,debenedetti2001supercooled,speedy2003kauzmann,berthier2011theoretical,welch2023cracking}. From a theoretical viewpoint, Kauzmann's key contribution was the proposal of a novel type of thermodynamic phase transition between an ergodic liquid and a non-ergodic, non-periodic solid. Developing accurate theories of the glass transition remains a fundamental open problem in classical statistical physics, with a growing body of work impacting various scientific fields (including protein folding~\cite{bryngelson1987spin}, biological processes~\cite{bi2016tissue}, computer science~\cite{monasson1999determining}, and machine learning~\cite{mehta2919ml}) where direct analogs of a Kauzmann transition were identified.
	
	The transition envisioned by Kauzmann has since been theoretically considerably refined, from the early work of Gibbs-diMarzio~\cite{gibbs1958nature} to its modern incarnation as a random first-order transition~\cite{lubchenko2007theory,RFOT_2012}. The patient elaboration of a consistent theoretical framework for the liquid-glass transition culminated in the last decade with the unification~\cite{parisi2020theory} of the statistical physics of liquids in large dimensions with the Parisi theory of disordered systems, first developed for spin glasses~\cite{mezard1987spin}. A random first-order phase transition with a vanishing equilibrium free energy difference between liquid and glass phases (re-interpreted as a vanishing configurational entropy $T S_c$, see Fig.~\ref{fig:kauzmann}a) is mathematically demonstrated in large dimensions, with known features~\cite{parisi2020theory,guiselin2022glass}. In particular, the crystalline phase plays no role in this analysis.
	
	There is a profound disconnect between the detailed understanding of the equilibrium phase transition between liquid and glass states at mean-field level, and the dramatic lack of solid results in two and three dimensions where this should apply. Several reasons explain this unsatisfactory situation. First, even modern experimental techniques fail to produce equilibrated samples down to $T_K$~\cite{ediger2017highly}. Second, analytic progress beyond mean-field is difficult, as this requires generalizing Parisi's theory to finite dimensions~\cite{charbonneau2017glass}. Third, despite notable recent progress~\cite{swap:ninarello2017, bolton2024ideal, fan2024ideal, PhysRevLett.134.128201,ghimenti2024irreversible}, numerical simulations have so far all failed to approach $T_K$ in bulk systems because equilibrium sampling of the complex energy landscape of glassy liquids at low enough temperatures is a computationally hard problem, even for finite systems of modest sizes~\cite{berthier2023modern}.  
	
	We overcome for the first time this longstanding sampling challenge in a realistic off-lattice glass model. We develop an original computational strategy that carefully integrates the complementary strengths of several advanced Monte Carlo techniques. We demonstrate that this innovative approach enables ergodic sampling of the configuration space of a two-dimensional glass model~\cite{jung2024normalizing} down to $T=0$ {for systems with up to 77 particles.} Our methodology goes far beyond previous attempts~\cite{parmar2023depleting,santen2000absence} that were restricted to temperatures significantly above the presumed Kauzmann transition. This methodological breakthrough allows us to fully characterize the thermodynamics of the system around and below the Kauzmann temperature, including its ground state and organization of low-lying excited states. Our configurational entropy and tailored free energy measurements~\cite{berthier2014novel} directly reveal, with no extrapolation required, the existence of a random first-order transition towards a low-temperature equilibrium glass phase characterized by a hierarchical organization of low-temperature states. By increasing the system size, the transition temperature shifts rapidly towards zero, while the properties of the transition remain unaltered. Our work offers the first glimpse of the equilibrium liquid-glass phase transition in finite dimensions. It should serve as a foundation for future simulation work using larger systems, three-dimensional, and molecular models~\cite{simon2025molecular} of glass-formers to fully illuminate the nature of the glass state of matter.

	\subsection*{Numerical strategy to achieve equilibration and ergodic sampling}
	
	\begin{figure*}	[t!]
		\includegraphics[width=0.95\linewidth]{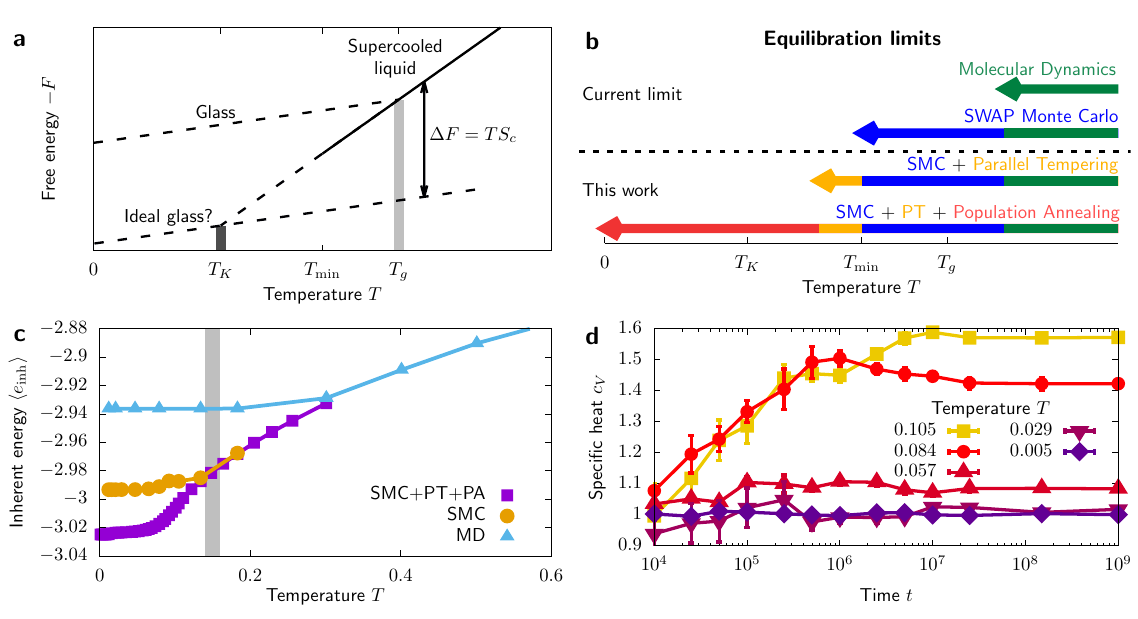}
		\caption{{Numerical strategy to study the equilibrium Kauzmann transition.}
			\textbf{a.} Location and definition of the Kauzmann transition. Above the   experimental glass transition temperature $T_g$ (light grey), the system is in a high entropy liquid state and becomes an arrested glass below $T_g$ due to lack of equilibration in conventional experiments. State-of-the-art simulations can probe equilibrium properties down to $T_{\rm min}$. 
			The Kauzmann temperature $T_K < T_{\rm min} < T_g$ denotes the temperature at which an equilibrium glass phase might form, with a free energy lower than that of the liquid. Above $T_K$, the free energy difference between liquid and glass, {$\Delta F = -(F_{\rm liq}-F_{\rm glass}) $}, defines the configurational entropy $T S_c$ that vanishes at $T_K$. 
			\textbf{b.} Performance and limits of numerical techniques. While conventional molecular dynamics simulations fail well above $T_g$, the swap Monte Carlo algorithm can reach below $T_g$. Here we combine parallel tempering (PT), population annealing (PA), and swap Monte Carlo (SMC) to enable equilibration down to $T\rightarrow 0,$ below the Kauzmann temperature $T_K.$
			\textbf{c.} Average inherent energy density $\langle e_{\rm inh}\rangle$ {at $N=77$} using different sampling techniques. MD and SMC respectively form non-equilibrium glasses below temperatures $T=0.3$ and $T=T_g$ when used to cool liquids {for $5\cdot 10^4\tau$}, whereas the combination SMC+PT+PA reaches deeper energy minima and eventually the ground state at $T=0$. 
			\textbf{d.} Demonstration of convergence for the sampling time $t $ dependent specific heat $c_V(t)$ at various temperatures $T.$ The long-time plateaus demonstrate ergodic equilibrium sampling. Our method reaches plateaus for all temperatures down to $T\rightarrow 0$, indicating equilibration.}
		\label{fig:kauzmann}
	\end{figure*}
	
	We developed a novel computational strategy to successfully sample the configuration space of a realistic glass-former across the bulk Kauzmann transition, see Fig.~\ref{fig:kauzmann}b. It is applied to the study of a ternary mixture of Lennard-Jones particles in two dimensions~\cite{jung2023predicting} (see Methods). We recently developed this computational model to fulfill multiple constraints relevant to the present study. (i) It is microscopically realistic (it resembles metallic glasses and previously studied glass models). (ii) It is robust against ordering (demixing, crystallisation). (iii) The swap Monte Carlo (MC) algorithm~\cite{swap:ninarello2017} alone provides a significant equilibration speedup compared to conventional molecular dynamics. (iv) This model was recently used to benchmark sampling algorithms~\cite{jung2024normalizing} and to devise the appropriate tools (based on specific heat measurements) that validate equilibration and ergodic sampling.  {We restrict ourselves to purely classical simulations, disregarding any quantum effects which may have an impact on the glass transition \cite{bray1980replica,novikov2013role}.}
	
	For reference, the onset temperature for slow dynamics is $T_o \approx 0.5$, the mode-coupling crossover (where conventional molecular dynamics becomes inefficient) is $T_{\rm mct} \approx 0.3$, {and the temperature at which relaxation is about 12 orders of magnitude slower than in the liquid is $T_g \approx 0.15$ (usually identified as the experimental glass transition temperature)}. Using swap MC, the system falls out of equilibrium near $T_{\rm min} \approx 0.12 < T_g$. Although it can be used below $T_g$, swap MC alone is insufficient to approach closely, or even cross, a putative thermodynamic Kauzmann singularity. We show in Fig.~\ref{fig:kauzmann}c the average \is energy $\langle e_{\rm inh} \rangle$ obtained after quenching to $T=0$ configurations obtained at temperature $T$~\cite{sastry1998signatures}. These data confirm that swap MC equilibrates more efficiently than molecular dynamics, but falls out of equilibrium near $T_{\rm min}$. 
	
	We discovered that a carefully designed combination of three known Monte Carlo techniques is exceptionally efficient at sampling the rugged configuration space at low $T$. To our knowledge, no other technique (or combination of techniques) has demonstrated such performance before. Our approach, sketched in Fig.~\ref{fig:kauzmann}b, is composed of two main steps. First, we combine swap MC with parallel tempering~\cite{swendsen1986replica,hukushima1996exchange} to produce large sets of $N_s = 10^6$ independent equilibrium configurations. For system sizes $N=44$, 55, 66, and 77, we can construct such equilibrium sets down to $T=0.105 < T_{\rm min}$. {The addition of PT improves over the performance of swap MC alone.} At this temperature, however, the system is still in the supercooled liquid phase. The second step uses the population of $N_s$ equilibrium configurations as starting points for the population annealing algorithm~\cite{hukushima2003population,machta2010population}. In this algorithm, the population is annealed to lower temperatures in a stepwise manner, with population resampling and reweighting steps performed at each $T$. Figure~\ref{fig:kauzmann}c demonstrates that population annealing yields much lower {averaged} \is energies, suggesting that it pushes the equilibration limit to much lower temperatures. {In itself, population annealing does not sample deeper energy minima, but the algorithm properly reweights and anneals the population to produce the correct equilibrium ensemble at lower temperatures.}
	
	Stringent tests are needed to decide whether ergodic sampling is achieved. To this end, we follow the time evolution of the specific heat, $c_V$, {by calculating the variance of the potential energy averaged up to a time $t$}. A correct determination of $c_V$ requires proper ergodic exploration of the configuration space. {This is a sensitive test as it demands that sampling is not restricted to a small part of configuration space~\cite{jung2024normalizing}.} In Fig.~\ref{fig:kauzmann}d we show, for selected temperatures, that the specific heat measured in the interval $0 \leq T \leq 0.105$ always converges to {a long-time plateau, which is a strong indication of equilibrium sampling. This first convergence test} was successful for all investigated sizes down to $T=0$, but failed at a finite temperature $T>0$ for larger $N$ values. This test sets the limit to the $N$-values reported in this work. See also Methods and supplementary information (SI) Figs.~S2 and S3 for several additional equilibration tests confirming these conclusions.
	
	In short, our innovative strategy carefully combining the known swap, parallel tempering, and population annealing Monte Carlo algorithms allows us to determine the equilibrium thermodynamic behavior of the system at any $T \geq 0$ for $N \leq 77$. In addition, the $T \to 0$ limit offers a direct determination of the ground state and opens a window into low-lying excitations, thus allowing a detailed analysis of the structure of the potential energy landscape deep in the glass phase for the first time.
	
	\subsection*{Configurational entropy}
	
	\begin{figure}[h]
		\includegraphics[width=0.95\linewidth]{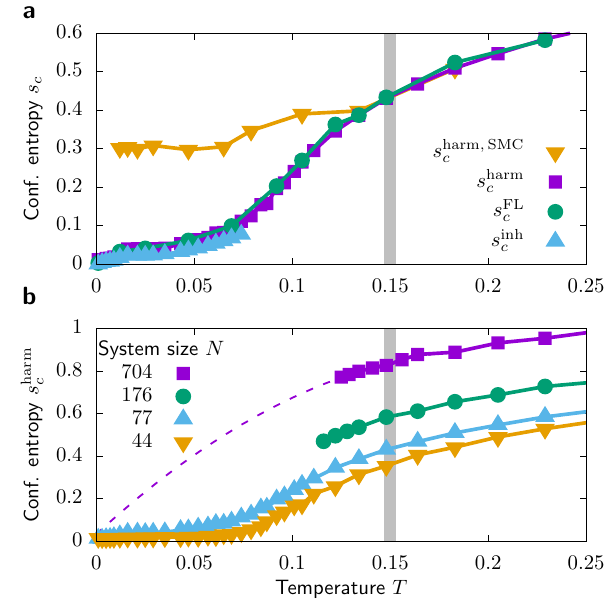}
		\caption{{Numerical measurements of the configurational entropy.} \textbf{a.} Configurational entropy for $N=77$ calculated using various techniques: expanding around \iss including anharmonic corrections ($s_c^{\rm harm}$), expanding around equilibrium states using the Frenkel-Ladd method ($s_c^{\rm FL}$), and Shannon entropy obtained from the distribution of \iss ($s_c^{\rm inh}$). When only using swap Monte Carlo, $s_c^{\rm harm,\,SMC}$ displays a non-equilibrium saturation at low $T$ when the system falls out of equilibrium. The agreement between the three methods is remarkable, and all equilibrium estimates indicate that the configurational entropy vanishes near $T=0$. The vertical light grey region indicates $T_g$.
			\textbf{b.} \Is configurational entropy $s_c^{\rm harm}$ for different system sizes $N$. Equilibrium is achieved down to $T=0$ for $N=44$ and 77. For $N = 176$ and 704, data are shown down to the lowest temperature for which equilibration is reached. The dotted line is an empirical extrapolation of the $N=704$ data using a quadratic polynomial fit.}
		\label{fig:s_c}
	\end{figure}
	
	In his original work~\cite{kauzmann1948nature}, Kauzmann defined a configurational entropy (or, rather, an `excess' one) as the difference between the supercooled liquid and crystal entropies. Extrapolation to temperatures below $T_g$ suggested the existence of what is now broadly called the Kauzmann temperature, as the moment where this excess entropy may vanish. Theoretical understanding, however, has significantly improved and configurational entropy must be defined independently of the crystal. What is physically relevant is the entropy difference with respect to amorphous glassy states~\cite{sciortino1999inherent,sastry2001relationship,berthier2014novel}.
	
	{In this more modern perspective there is no unique partitioning of phase space into vibrations and configurations. Therefore, several distinct computational methods were developed to estimate such a configurational entropy from simulations~\cite{sciortino1999inherent,berthier2019confentropy}. A first line of research is deeply rooted in the potential energy landscape view of glasses~\cite{stillinger1988supercooled,sciortino1999inherent,sastry2001relationship,berthier2019confentropy}, and we shall first explore this route. In this view, the Kauzmann transition is re-interpreted as a vanishing of the inherent structure configurational entropy.}
	
	{Over the last 40 years, a second line of research has demonstrated that another formulation of a Kauzmann transition is generically found in theoretical models when considering instead the temperature evolution of the free-energy landscape \cite{derrida1980random,gross1984simplest,singh1985hard,kirkpatrick1987dynamics,kirkpatrick1987p,franz1997phase,mezard1999first,biroli2001lattice,kurchan2013exact}. These studies demonstrate, as anticipated by Stillinger himself \cite{stillinger1988supercooled}, that enumerating potential energy minima cannot lead to the transition as envisioned by Kauzmann, whereas free-energy measurements can. In this second approach, the Kauzmann transition happens when glass and liquid free energies become equal.}
	
	{Our numerical study is capable of equilibrating supercooled liquids down to $T=0$, and it should thus be able to distinguish between these different types of partitioning and concrete measurements, potentially revealing both the vanishing of the configurational entropy associated with localized defects at $T=0$, as envisioned by Stillinger, and, at the same time, the presence of a finite temperature free energy singularity characterizing instead the evolution of the free-energy landscape.}
	
	In Fig.~\ref{fig:s_c}a, we report three independent estimates of the configurational entropy per particle, $s_c$, for $N = 77$ {based on inherent structures.} The first estimate uses an expansion around inherent structure energy minima~\cite{sciortino1999inherent,Sciortino_2005} and the second one relies on a thermodynamic integration from a reference liquid configuration~\cite{ozawa2018configurational}, in close analogy to the Frenkel-Ladd method to determine the free energy of crystals~\cite{frenkel1984new}. The third approach uses a Shannon entropy associated with the distribution of \iss\cite{parmar2023depleting}. See Methods for precise definitions. 
	
	Remarkably, these distinct methodologies are in excellent agreement, with small discrepancies that are within the numerical accuracy, see Fig.~\ref{fig:s_c}a. To confirm further that equilibration has been reached at all temperatures, we performed these configurational entropy measurements using swap MC alone. In that case, $s_c$ decreases to a finite plateau when $T < 0.12$, due to non-ergodic sampling, see Fig.~\ref{fig:s_c}a. The consistent finding across three independent methods that $s_c(T=0)=0$ is by no means trivial because in each method $s_c$ is the superposition of several large terms that can only cancel if statistical uncertainties are small and the system properly equilibrated. The physical interpretation is that, when equilibrium can be maintained in the limit $T \to 0$, the system ends in a non-degenerate ground state that we successfully identify.
	
	As temperature decreases the configurational entropy decreases sharply with a pronounced drop below $T_g$. This precipitous drop mimics the key experimental observation noted by Kauzmann~\cite{kauzmann1948nature}. However, because we can follow $s_c$ all the way to $T=0$ we can directly demonstrate that $s_c$ only vanishes at $T=0$. {Therefore, in our two-dimensional system, the system-size dependence of the configurational entropy $s_c(T)$ shows no sign of a finite temperature entropy crisis using such estimates.} This is reasonable since the \is picture of the thermodynamics of glasses~\cite{PhysRevA.25.978} on which these estimates are based was developed much before the nature of random first-order transitions based on free energies was fully understood.
	
	In Fig.~\ref{fig:s_c}b, we provide a broader perspective on the behavior of the configurational entropy by studying its dependence on $N$ over a wide range. Our measurements from $N=44$ to $N=77$ are all consistent with a smooth evolution and an entropy crisis taking place at $T=0$ exactly. For larger systems, $N=176$ and $N=704$ (where the bulk limit is nearly reached) we can no longer equilibrate the system at very low temperatures. The dashed line through the $N=704$ data is a hypothetical extrapolation used earlier~\cite{berthier2019zero} to estimate the entropy to vanishing temperatures.
	
	The behavior reported in Fig.~\ref{fig:s_c}a is qualitatively consistent with the picture proposed by Stillinger who imagined very low-temperature configurations as an amorphous ground state decorated by {localized} excitations~\cite{stillinger1988supercooled}. Our numerical results are in line with this view (see our study of low-lying excitations below), and {confirm that $s_c$ does not vanish at a finite temperature when it is defined from inherent structures.}
	
	{Stillinger further interpreted the smooth approach of the inherent structure configurational entropy to zero at $T=0$ as demonstrating more generally the absence of any kind of Kauzmann transition \cite{stillinger1988supercooled}. The results below paint a very different picture, and show that Stillinger's conclusions stemmed from the inability of the inherent structure approach to capture the evolution of the free-energy landscape. Because energy minima do not distinguish between localised and collective excitations, Stillinger's entropy is blind to the presence of a thermodynamic singularity in the free energy landscape. A different analysis is needed to investigate whether finite-size systems exhibit a free energy-based Kauzmann transition into an ideal glass phase, as presented in the next section.}
	
	\subsection*{Free energy measurements and the Kauzmann transition}
	
	\begin{figure}[h]
		\includegraphics[width=0.95\linewidth]{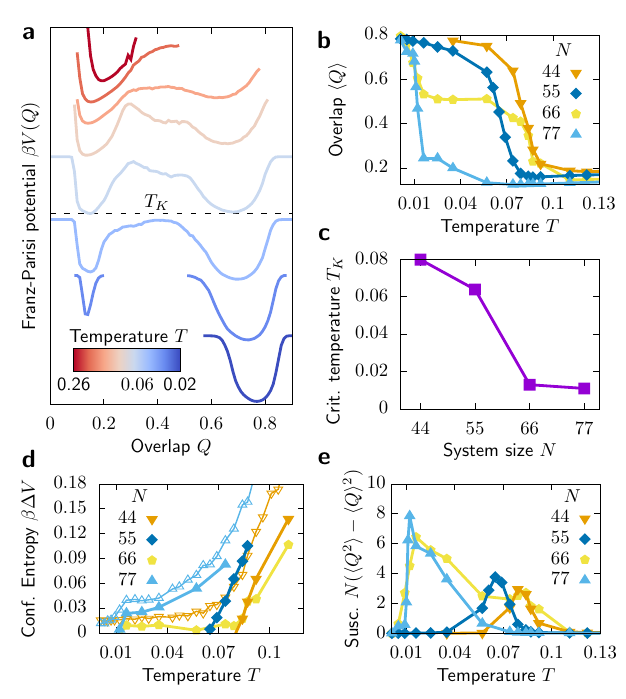}
		\caption{{First-order equilibrium Kauzmann transition revealed by the evolution of the Franz-Parisi free energy.} 
			\textbf{a.} The Franz-Parisi potential $\beta V(Q)$ as a function of the overlap order parameter $Q,$ between independent pairs of equilibrium configurations for $N=55$. The potential is vertically shifted arbitrarily for visibility. For $T$ far above $T_K$ the potential has a single minimum at small $Q$ corresponding to the liquid phase. A secondary minimum appears at large $Q$ above $T_K$, corresponding to a metastable glass. At $T=T_K$, both minima have the same free energy. Below $T<T_K$, the large $Q$ glass minimum dominates. The evolution of the free energy is reminiscent of conventional first-order phase transitions.    
			\textbf{b.} Temperature-dependent average overlap $\langle Q \rangle$ for different system sizes $N$ reveals a sharp change from small to large at a size-dependent transition temperature. $N = 66$ is an outlier because it develops a quasi-ordered structure with special symmetries, which distorts the overlap statistics as discussed in the SI.  
			\textbf{c.} Evolution of the critical Kauzmann temperature $T_K$ with system size, suggesting that $T_K \rightarrow 0$ for large $N$. This is consistent with the entropy extrapolation for $N=704$ in Fig.~\ref{fig:s_c}b. \textbf{d.} Temperature-dependent configurational entropy $s_c = \beta (V(Q_{\rm glass}) - V(Q_{\rm liq}))$ as extracted from the Franz-Parisi potential {(full symbols), compared to the results shown in Fig.~\ref{fig:s_c}b (open symbols)}. \textbf{e.} Susceptibility of the order parameter calculated via the fluctuations of the overlap $Q.$ The susceptibility reveals a pronounced peak at $T_K$ which grows with $N.$}
		\label{fig:Franz_Parisi}
	\end{figure}
	
	{To detect a Kauzmann transition from free energies, we employ the formalism first introduced by Franz and Parisi in the context of mean-field glass models~\cite{franz1997phase}, and extensively used in recent numerical studies of finite-dimensional supercooled liquids~\cite{berthier2013overlap,berthier2014novel,guiselin2022statistical}.} We study an effective Landau free energy, $V(Q)$, expressed as a function of the overlap $Q$. The overlap quantifies the similarity of the collective density fields in two independent configurations (see Methods). Because $Q$ is insensitive to localized excitations, it is the correct order parameter to detect the liquid-glass Kauzmann transition. The function $V(Q)$, known as the Franz-Parisi potential, is the free energy cost to observe the value $Q$ of the overlap with respect to a fixed reference equilibrium configuration.
	
	In this framework, the Kauzmann transition resembles a conventional first-order phase transition associated to a jump in the order parameter $Q$~\cite{guiselin2022glass}. {Consistent with previous work, we will identify the critical temperature of this transition as $T_K$ \cite{mezard1999thermodynamics}.} In the liquid above $T_K$, the global minimum of $V(Q)$ is at low overlap, $Q \approx Q_{\rm liq}$, indicative of a liquid phase in which the system accesses many different configurations with small mutual overlap. In the glass below $T_K$, $V(Q)$ displays a different global minimum at a much larger value, $Q \approx Q_{\rm glass}$, revealing that the equilibrium glass only samples a limited number of amorphous states. In this regime, there is a finite probability to remain close to a reference configuration, thus leading to a large value of $Q$. The overlap jumps discontinuously from $Q_{\rm liq}$ to $Q_{\rm glass}$ at $T_K$, where the free energy difference $\Delta V = V(Q_{\rm glass}) - V(Q_{\rm liq})$ vanishes. In contrast to the conventional estimates of the configurational entropy shown in Fig~\ref{fig:s_c}, the alternative definition $T s_c = \Delta V$ can lead to a genuine entropy crisis~\cite{berthier2014novel}, with $\Delta V \to 0$ at a finite temperature.   
	
	To determine $V(Q)$, we measure the mutual overlap $Q_{mn}$ between two independent equilibrium configurations $m$ and $n$. Given two independent sets of $N_s$ equilibrium structures at temperature $T$ we evaluate the quenched Franz-Parisi potential as: 
	\begin{equation}
	\beta V(Q) = - \frac{1}{N N_s}\sum_{m=1}^{N_s}  \ln \left( \sum_{n=1}^{N_s} \delta(Q-Q_{m n}) \right),
	\end{equation}
	with $\beta = 1/T$.
	Unlike previous approaches, which relied on biased Monte Carlo techniques to probe rare fluctuations of $Q$ much above $T_K$~\cite{berthier2013overlap}, we directly construct $V(Q)$ from unbiased equilibrium sampling at any $T$. This is possible because the system is equilibrated near $T_K$, where the relevant fluctuations of $Q$ become typical. Biasing techniques towards rare fluctuations of $Q$ could inform about tails of the distribution function of the overlap, the Parisi function $P(Q) \sim \exp( - \beta N V(Q))$.
	
	Measuring the Franz-Parisi potential down to $T=0$ reveals our main physical result, see Fig.~\ref{fig:Franz_Parisi}. A Kauzmann transition from the liquid (with a small average overlap) to the ideal glass phase (with a large average overlap) takes place when the glass free energy becomes lower than that of the liquid, as directly observed in Fig.~\ref{fig:Franz_Parisi}a. At temperatures much larger than $T_K$, the system exhibits low-overlap configurations with \(Q \simeq  Q_{\rm liq} \), characteristic of the liquid state. As the temperature decreases, a second local minimum appears at high overlap \( Q_{\rm glass} \), becoming degenerate with the liquid minimum at a well-defined temperature that we identify as the Kauzmann temperature: \( T_K = 0.057 \) for \( N = 55 \). Below that temperature $T_K(N)$, the system enters the low-temperature equilibrium glass, characterized by the dominance of the high-overlap minimum. Correspondingly, the average overlap jumps abruptly from low to large value in a narrow temperature regime near $T_K$, as shown in Fig.~\ref{fig:Franz_Parisi}b. Besides, the temperature evolution of $s_c=\beta \Delta V$ in Fig.~\ref{fig:Franz_Parisi}d directly reveals that it continuously vanishes at a well-defined Kauzmann temperature. {At high $T$, $\beta \Delta V$ is similar to $s_c$ as extracted in Fig.~\ref{fig:s_c}b up to a constant shift, which emerges since many different inherent structures can be assigned to the same metastable state. However, at low $T$ both measures differ qualitatively, highlighting the different nature of the inherent structure formalism, which can be dominated by localized excitations.} We emphasize that all these low-temperature free energy and overlap measurements across $T_K$ are performed in fully equilibrated conditions.  
	
	Phase transitions are necessarily rounded in finite systems and the free energy $V(Q)$ is possibly not convex~\cite{guiselin2022statistical}. To begin to understand finite-size corrections, we studied four system sizes. The range of sizes covered is admittedly modest, but we decided to restrict to $N$ values for which ergodicity could be convincingly established, and statistical uncertainties in all reported quantities were small. Yet, the reported evolution with $N$ provides significant insight as we now detail.
	
	The evolution of both the average overlap and of \( \beta \Delta V \) indicates a systematic shift of \( T_K(N) \) to lower temperatures with increasing \( N \), see Fig.~\ref{fig:Franz_Parisi}c. These data show that $T_K(N)$ rapidly shifts towards zero as $N$ increases. Additionally, the overlap fluctuations $\chi = N [\langle Q^2 \rangle - \langle Q\rangle^2 ]$ display a peak near $T \approx T_K(N)$, with an amplitude that increases rapidly with $N$, see Fig.~\ref{fig:Franz_Parisi}e. Although the accessible system sizes do not permit a quantitative finite-size scaling analysis, they indicate (i) \( T_K \simeq 0.01 \) for the largest $N$ studied, which is substantially lower than the glass transition temperature \( T_g \approx 0.15 \), and conspicuously close to $T_K=0$, (ii) a trend of $T_K(N)$ decreasing with $N$. The most natural interpretation of our data is that $T_K(N)$ would continue to approach zero if larger systems could be studied. We conclude that our numerical results are most consistent with a theoretical scenario in which the Kauzmann transition occurs at \( T_K = 0 \) in the large $N$ limit. We hope that future studies can strengthen this conclusion by investigating larger system sizes. 
	
	\subsection*{Amorphous ground states and hierarchy of low-lying excitations} 
	
	\begin{figure*}
		\includegraphics[scale=0.63]{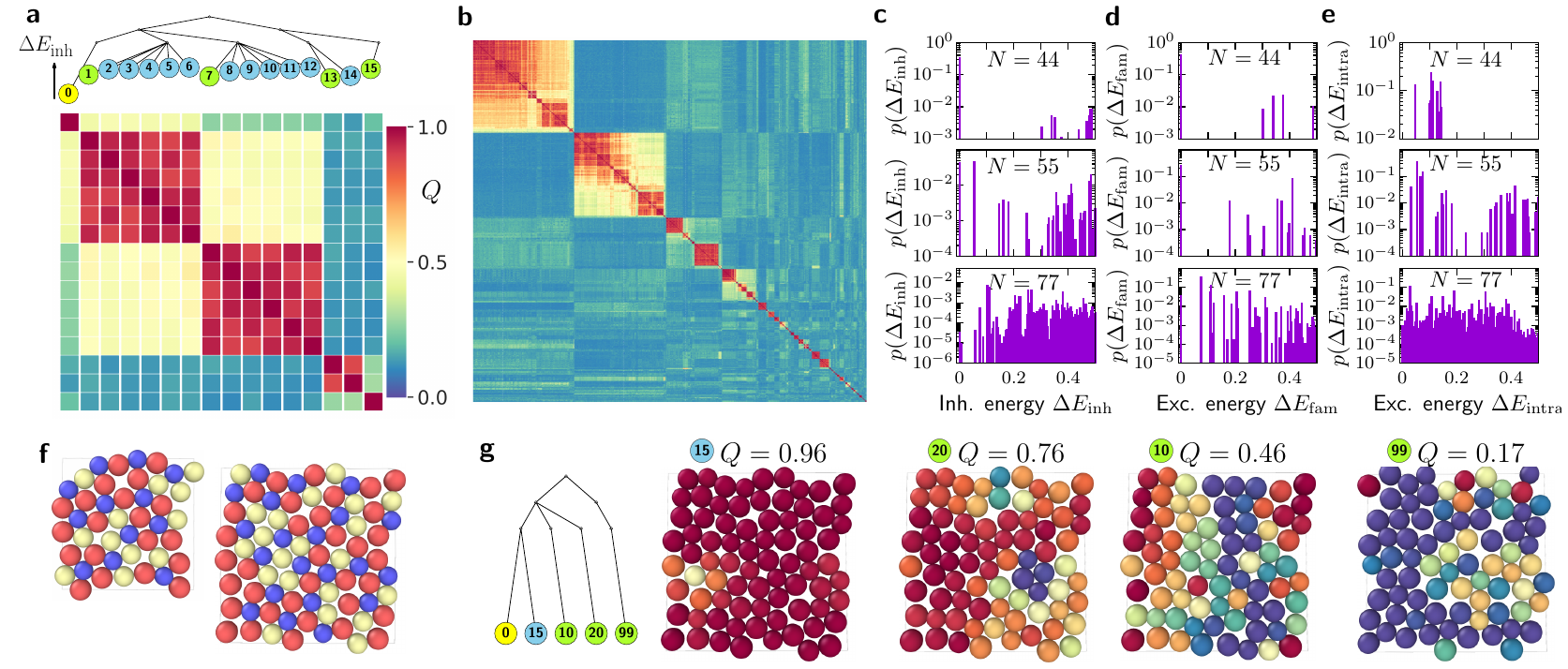}
		\caption{{Hierarchical organization of low energy states.} 
			\textbf{a.} Tree and matrix representation of the bottom of the potential energy landscape for $N=44$. States with mutual overlaps $Q>0.8$ are grouped into families. The family minimum is shown with a green circle and the ground state with a yellow circle, excited states as blue circles. In the tree, the x-axis is arbitrary, the y-axis is the inherent energy. Using the same ordering of states as in the tree structure, the matrix represents the value of the mutual overlap between all considered pairs of \iss, shown by the color bar, and reveals the hierarchical organisation of the bottom of the energy landscape.  
			\textbf{b.} Same as in \textbf{a} for $N=77$.  
			\textbf{c.} Probability of finding states with inherent energy $\Delta E_{\rm inh}$ for various system sizes $N= 44$, 55, and 77.  
			\textbf{d.} Probability of inter-family excitation energies $\Delta E_{\rm fam}$ for the same $N$ values.  
			\textbf{e.} Probability of intra-family excitation energies $\Delta E_{\rm intra}$ for the same $N$ values. The proliferation of both distinct minima and localized excitations for increasing $N$ is obvious.  
			\textbf{f.} Snapshot of the ground states for $N=44$ and $N=77$. Both structures are fully amorphous without any signature of crystallization. The color encodes the particle type. 
			\textbf{g.} Excerpt of the $N=77$ tree, and corresponding snapshots of the selected excited states showing various degree of overlap $Q$ with the ground state. The color encodes the local overlap for each particle using the same color code as in \textbf{a}.}
		\label{fig:States}
	\end{figure*}
	
	We now study the lowest energy configurations to investigate whether the Kauzmann transition is associated with a rich and complex structure of the free energy landscape. For all investigated $N$, we obtain global ground states that are completely amorphous, as illustrated in Fig.~\ref{fig:States}f for $N=44$ and $77$. This confirms that crystallization is efficiently suppressed. We cannot exclude that a periodic structure with a very large unit cell eventually emerges for much larger $N$. 
	
	To explore the organization of low-energy minima, we classify all identified \iss into families based on structural similarity between them. The structure of the Franz-Parisi potential indicates that pairs of configurations are either very similar or markedly distinct. We define a family as a set of \iss with mutual overlap $Q > 0.8$. This threshold value was chosen based on the results shown in Fig.~\ref{fig:Franz_Parisi}.
	
	Each \is is characterized by two excitation energies: the excitation energy relative to the global ground state, $\Delta E_{\rm inh} = E_{\rm inh} - E_{\rm gs}$, and an intra-family excitation energy,  $\Delta E_{\rm intra} = E_{\rm inh} - E_{\rm fam}$, where  $E_{\rm fam}$ denotes the minimum energy within the same family. We can also define the position of a given family relative to the ground state, $\Delta E_{\rm fam} = E_{\rm fam} - E_{\rm gs}$. By construction, one has $\Delta E_{\rm inh} = \Delta E_{\rm intra} + \Delta E_{\rm fam}$. This decomposition is useful, as we can disentangle intra-state ({localized}) excitations from inter-states ({collective}) excitations directly in our simulations.
	
	We focused our analysis on low-lying minima, specifically those with excitation energies $\Delta E_{\rm inh} \ll |E_{\rm gs}|/N \approx 3.0$, as illustrated in Fig.~\ref{fig:States}c. The probability distributions of \iss as a function of the excitation energies defined above are shown in Fig.~\ref{fig:States}c, and their decompositions into localized and collective excitations in Figs.~\ref{fig:States}d,e for three system sizes. We observe a proliferation of low-lying states as $N$ increases, with a nearly equivalent contribution from both types of excitation energies. Thus, the bottom of the potential energy landscape is composed of several distinct families, themselves decorated by localized excitations, with a complexity that increases rapidly with $N$. 
	
	To visualize the organization of states and localized excitations, we compute the matrix of mutual overlap between low-lying states~\cite{berthier2019gardner} alongside the corresponding clustering trees in Figs.~\ref{fig:States}a,b. In the tree representation, each circle represents an \is, with the vertical position reflecting its inherent energy, while connections in the tree are based on mutual overlaps and constructed via a clustering algorithm (see Methods). The resulting structure reveals a clear hierarchy: \iss are grouped into families, and families in turn cluster into higher-order families of families. This organization reveals a complex, hierarchical structure at the bottom of the energy landscape, even after localized excitations have been coarse-grained. The complex organization of collective excitations, and its evolution with temperature, gives rise to the Kauzmann transition {reported in Fig.~\ref{fig:Franz_Parisi}c. Instead, the existence of localized excitations within families implies that the inherent structure entropy does not vanish until $T=0$.}
	
	Finally, a real-space view of the hierarchical phase-space organization is provided in Fig.~\ref{fig:States}g. We provide one snapshot of a localized excitation within the ground-state family, two snapshots of collective excitations from distinct families with a finite overlap with the ground state, and one snapshot from a structurally distinct family. These snapshots reveal that intra-family excitations correspond to localized rearrangements, while transitions between different families involve increasingly collective up to system-spanning reorganizations. These observations are in agreement with the bimodal overlap distribution obtained from the Franz-Parisi free energy potential. Once again all these structures are fully amorphous. 
	
	\subsection*{Discussion}
	
	Because we can directly access all thermodynamic quantities down to arbitrarily low temperatures, our results clarify the behavior of glass-forming materials when equilibrated at temperatures that have so far remained inaccessible to both experiments and simulations. 
	
	First, our analysis demonstrates that the existence of the crystal phase can be, and should be when possible, separated from the discussion of the Kauzmann transition. Second, we find that localized excitations, invoked by Stillinger~\cite{stillinger1988supercooled,sastry2004numerical}, indeed proliferate with system size. This is expected since their energy cost is not extensive. However, the Franz-Parisi free energy naturally coarse-grains these excitations and allows us to probe the properties of collective states. The finite-size Kauzmann transition described in Fig.~\ref{fig:Franz_Parisi} characterizes the temperature evolution of those states, and is unaffected by localized defects. Our numerical results directly confirm Stillinger's findings regarding localized excitations (Fig.~\ref{fig:States}) and the smooth vanishing of the \is configurational entropy (Fig.~\ref{fig:s_c}), but our study of the free energy associated to overlap fluctuations establishes that a finite temperature Kauzmann transition can be observed (Fig.~\ref{fig:Franz_Parisi}), therefore combining some of the original Stillinger's ideas with more recent theoretical results regarding the Kauzmann transition.
	
	Our findings provide compelling evidence that the landscape of low-lying states is intrinsically complex and organized in a hierarchical manner, as envisioned in Parisi's theory of disordered systems~\cite{parisi2020theory,berthier2019gardner}. We uncover a very similar picture with clusters of low-energy configurations which are nested within one another, forming a multi-level hierarchy. The bottom of the energy landscape is populated by minima whose energy differences are much smaller than the energy per particle. Yet, these minima correspond to completely different configurations in real space, with very small mutual overlap, thus reflecting global, system-wide rearrangements.  Considerable additional numerical work is required to characterize all details of the Parisi function $P(Q)$ and its dependence on system size, in order to connect these results to a possible class of replica symmetry–breaking.
	
	The most natural theoretical interpretation of our results is that a non-trivial Kauzmann transition, based on free-energy measurements, occurs at $T_K=0$ in our system when $N$ is large. While this hypothesis was suggested before using temperature extrapolations~\cite{berthier2019zero,parmar2023depleting}, this conclusion is now supported by direct equilibrium measurements down to $T = 0$ for finite $N$ systems. Testing whether the same picture holds for more complex models and three-dimensional systems is an exciting and important challenge, for which our work lays the conceptual and methodological foundation.
	
	Our conclusions should also motivate theoretical work to incorporate finite-dimensional fluctuations into the framework of random first-order transitions. In fact, analogies to random field Ising models~\cite{stevenson2008constructing,biroli2018random} do not suggest the possibility of a zero-temperature critical point in two-dimensional glasses. However, the temperature dependence of the effective couplings (information that can now be accessed by combining our results with the approach developed in Ref.~\cite{stevenson2008constructing}) may instead reveal a different scenario. We hope that our solid numerical results will provide useful guidance and motivation for future work.
	
	The Kauzmann transition has long been seen as being an abstract concept of paramount conceptual importance (it questions the existence of a disordered state of matter \cite{anderson1995through}), but perhaps of little practical relevance since it concerns a temperature regime that cannot be reached using conventional cooling methods or numerical techniques. Yet recent years have seen dramatic progress in the computational~\cite{berthier2023modern} and experimental~\cite{ediger2017highly,Royall_2018,monnier2021reaching} realms. Our work connects, in particular, to recent experimental advances preparing ultrastable glasses by physical vapor deposition. This leads to amorphous films whose thermodynamic stability approaches very nearly that expected near the Kauzmann temperature~\cite{ediger2017highly,rodriguez2022ultrastable,beasley2019vapor}. Although these materials are produced through non-standard protocols, they provide an increasingly close experimental realization of the ideal-glass regime. Our approach therefore offers a natural framework for interpreting these experiments from an equilibrium perspective, while providing a benchmark for future studies of the thermodynamics, dynamics, and mechanical properties of such deeply equilibrated amorphous solids possibly leading to experimental demonstrations, hopefully in the coming years, of the ideal glass state of matter.
	
	\section*{Data availability statement}
	
	The low-temperature configurations and the code to extract configurational entropies and Franz-Parisi free energies have been uploaded on zenodo (\url{https://zenodo.org/records/21774618}).
	
		\section*{Acknowledgments}
	
	MO thanks the support by MIAI@Grenoble Alpes and the Agence Nationale de la Recherche under France 2030 with the reference ANR-23-IACL-0006). This work was also supported by a grant from the Simons Foundation (LB: \#454933, GB: \#454935). GB acknowledges funding from the French government under management of Agence Nationale de la Recherche as part of the “Investissements d’avenir” program, reference ANR-19-P3IA-0001 (PRAIRIE 3IA Institute). LB acknowledges the support of the French Agence Nationale de la Recherche (ANR), under grants ANR-20-CE30-0031 (project THEMA) and ANR-24-CE30-0442 (project GLASSGO).
	
	\bibliography{library_local.bib}

	\FloatBarrier
	
	\setcounter{equation}{0}
	\setcounter{figure}{0}
	\setcounter{table}{0}
	\renewcommand{\theequation}{S\arabic{equation}}
	\renewcommand{\thefigure}{S\arabic{figure}}
	\renewcommand{\bibnumfmt}[1]{[S#1]}
	
	\newpage
	
	\section*{Methods}
	
	\section{Two-dimensional atomistic glass-former}
	\label{sec_met:model}
	
	The atomistic glass-former simulated in this work is very similar to the model introduced in Ref.~\cite{jung2023predicting}. The model corresponds to a modified Kob-Andersen mixture in two-dimensions (KA2D) in which particles interact via a Lennard-Jones potential,
	\begin{equation}\label{eq:LJ}
	V_{\alpha \beta}(r_{ij}) = \begin{cases}
	4 \epsilon_{\alpha \beta} \left[\left(\frac{\sigma_{\alpha \beta}}{r_{ij}}\right)^{12} - \left(\frac{\sigma_{\alpha \beta}}{r_{ij}}\right)^6 + C_0 \right. & \\  \,\left. + C_2\left( \frac{r_{ij}}{\sigma_{\alpha \beta}}\right)^{2}+ C_4\left( \frac{r_{ij}}{\sigma_{\alpha \beta}}\right)^{4}  \right] & r_{ij}<r^\text{cut}_{\alpha \beta}\\
	0 & \text{otherwise},
	\end{cases}
	\nonumber
	\end{equation}
	for particle distance $r_{ij}= |\bm{r}_i - \bm{r}_j|$, where $\bm{r}_i$ is the position of the $i$-th particle.
	The ternary KA2D model consists of three types $\alpha,\beta = \{1,2,3\}$ where types 1 and 2 interact via the usual Kob-Andersen non-additive interactions, $\epsilon_{11}=1.0$, $\epsilon_{12}=1.5$, $\epsilon_{22}=0.5$ and $\sigma_{11}=1.0$, $\sigma_{12}=0.8$, $\sigma_{22}=0.88.$ Additionally, we introduce a third species~\cite{Berthier2020} with interaction parameters $\epsilon_{13}=0.75$, $\epsilon_{23}=1.5$, $\epsilon_{33}=0.75$ and $\sigma_{13}=0.9$, $\sigma_{23}=0.8$, $\sigma_{33}=0.94.$ The cutoff depends on the particle type $r^\text{cut}_{\alpha \beta} = 2.5\sigma_{\alpha \beta}$ and we define the constants $C_0= 0.04049023795$, $C_2= -0.00970155098$ and $C_4= 0.00062012616$ to make the potential continuous up to the second derivative. Differently from Ref.~\cite{jung2023predicting} we slightly change the type composition to a ratio 5:3:3.
	The system parameters were empirically adjusted to avoid any signatures of local crystallization, and indeed no crystalline structure was found in any of the simulation runs performed. Results are reported in reduced Lennard-Jones units defined by $\epsilon_{11}$ (energy scale), $\sigma_{11}$ (length scale) and $\tau=\sigma_{11}\sqrt{m/\epsilon_{11}}$ (time scale), with mass $m=1$ for all types.
	
	The particles evolve in a square box of dimensions $L_x=L_y$ with constant particles density $\rho = N / (L_xL_y) = 1.19$, where $N$ is the number of particles. We use $N=44$, 55, 66, 77, 176, and $704$ in this study. The linear box sizes are therefore $L_x(N=44)=6.08$, $L_x(N=55)=6.8$, $L_x(N=66)=7.44$, $L_x(N=77)=8.036$, $L_x(N=176)=12.16$ and $L_x(N=704)=24.32$.
	
	The relaxation dynamics of the KA2D model using MD dynamics is shown in SI Fig.~S1. We find the expected relaxation dynamics as featured by the incoherent scattering function (ISF) for both $N=704$ (SI Fig.~S1a) and $N=77$  (SI Fig.~S1b). We note that the effects of Mermin–Wagner fluctuations~\cite{shiba2016unveiling} are small and negligible at these system sizes~\cite{jung2023predicting}. In particular, the ISF shows a decay towards a plateau, highlighting vibrational dynamics in cages followed by a long-time structural relaxation. Importantly, the $N=77$ system behaves dynamically as a typical bulk supercooled liquid with negligible finite size effects~\cite{buchner1999potential}. As expected, the relaxation times for $N=77$ are slightly higher than for $N=704$ due to the small box size~\cite{kim2000apparent,berthier2012finite}, which includes additional constraints (compare SI Fig.~S1c and d) but both systems show an apparent Arrhenius behavior, $\tau_\alpha^{\rm ISF}(T) \propto \exp(6.1 / T)$, for small temperature. The glass transition temperature $T_g$ can be identified as, $\tau_\alpha^{\rm ISF}(T_g)= 10^{12},$ yielding $T_g = T_g(N=704) = 0.147$ and $T_g(N=77)=0.156.$ We will report the former value throughout this work since it is the value we expect in the thermodynamic limit, but the difference is not very important, considering that we have mainly focused on temperatures $T<0.1$ in this manuscript.
	
	\section{\Is energy}
	
Given a set of $N_c$ equilibrium configurations, we can extract ensemble averages. In particular, we calculate the ensemble-averaged inherent energy density, $\langle e_{\rm inh} \rangle = \langle E_{\rm inh} \rangle / N$. For each configuration $m$ with particle coordinates $\bm{R}_m=(\bm{r}_{m,1},\bm{r}_{m,2},...,\bm{r}_{m,N})$ we determine the nearest \is, $\bm{R}_{\text{inh},m}.$ This is achieved by performing steepest descent with a maximal step size per particle $\Delta r= 0.01$ to minimize the potential energy $E_{\rm pot}.$ Finally, we define $E_\text{inh}= \sum^\prime_{i,j} V(r_{\text{inh},ij})/2$, as the potential energy of the energy minimum.
	
	\section{Criteria for equilibration and sampling}
	
	\label{sec_met:time_dep_cv}
	
Based on a time-dependent specific heat $c_V(t)$ we have recently established a reliable and independent measure to test whether an ensemble of supercooled liquid configurations represents an equilibrium ensemble or not~\cite{jung2024normalizing}. Here, the `time' $t=M \tau$ corresponds to the number $M$ of configurations used for averaging, multiplied by the relaxation time $\tau$ used between the production of two independent configurations. To define this measure we introduce the average, $\langle X \rangle_{M} = M^{-1}\sum_{m=1}^{M} X_m$ of an observable $X_m$ evaluated for configuration $m$. Using this definition, we obtain the potential energy, $\langle E_{\rm pot} \rangle_{M} $ and specific heat at temperature $T$, $c_V(t=M \tau) = (\langle E_{\rm pot}^2 \rangle_{M} - \langle E_{\rm pot} \rangle_{M}^2 ) / (NT^2).$ 

If a set of $N_c$ configurations represents an equilibrium ensemble, we expect $c_V(t)$ to increase from $c_V(t)=1,$ corresponding, by equipartition in a two-dimensional system, to thermal vibrations close to an arrested structure, and reach a long-time plateau $c_V=c_V(t\rightarrow \infty)$ that corresponds to the equilibrium value of the specific heat. In the case that $c_V(t=N_c \tau)$ has not yet reached a plateau, this clearly indicates that the set is not sufficiently equilibrated~\cite{jung2024normalizing}. For $N=77$, which was by far the most difficult system size to equilibrate (by comparison with smaller $N$ values) we observe from Fig.~\ref{fig:kauzmann}d in the main text, that $c_V(t)$ attains a plateau at all investigated temperatures. This provides compelling evidence that we can create equilibrium ensembles down to very low temperatures. This test fails whenever we do not use the combination of Monte Carlo techniques described in the main text.

	\section{Numerical techniques enabling equilibration down to zero temperature}
	
	\label{sec_met:numerical_techniques}
	
	Equilibration is achieved by the unique combination of three different Monte-Carlo techniques: swap Monte-Carlo (SMC), parallel tempering (PT) and population annealing (PA).
	
	\subsection{Swap Monte Carlo algorithm}
	
	\label{sec:SMC}
	
	SMC is based on the idea of accelerating equilibration by swapping the positions of pairs of particles of different types~\cite{swap:GAZZILLO1989,swap:ninarello2017,swap:Berthier2019}. The model introduced in the previous section was constructed to enable the best usage of SMC via the insertion of the intermediate type, which massively increases the swap probability~\cite{Berthier2020}.
	
	First, we create configurations with the composition defined above. Then, we use SMC in combination with molecular dynamics (MD) simulations for short thermalization runs of approximately $N_{t,\text{th}}=10^8$ time steps at $T=0.148.$ The molecular dynamics time step is set to $\Delta t = 0.005$ and the system is thermalized using a Nosé–Hoover thermostat~\cite{NH}. Every 10 MD steps we include a phase of SMC moves. In each phase we perform $2N$ swap attempts, by randomly choosing two particles of different type, swapping their position, and accepting the attempt via the Metropolis acceptance criterion~\cite{swap:Berthier2019}. All simulations are performed using LAMMPS~\cite{LAMMPS}. 
	
	\subsection{Parallel tempering}
	
	The thermalized configurations at $T=0.148$ by SMC are then used as input for PT simulations. The main idea behind PT is to perform parallelized SMC simulations of several replicas of the system~\cite{hukushima1996exchange}. Each replica evolves at a different temperature. Regularly, exchange events are attempted in which two replicas exchange their current configuration, which is accepted via the Metropolis acceptance criterion. If successful, configurations will travel in temperature space by visiting different replicas~\cite{swendsen1986replica,yamamoto2000PT}. Configurations originally simulated at a low temperature can thus randomly be exchanged between replicas, relax at high temperature, and possibly travel back to low temperature. This procedure has been shown to accelerate sampling in many instances.
	
	We perform simulations of 8 replicas in parallel (9 for $N=77$) at temperatures $T=0.256$, 0.229, 0.205, 0.183, 0.164, 0.148, 0.134, and $0.122$ (and $T=0.111$ for $N=77$). Exchange events are attempted every $10^3$ MD steps. The microscopic dynamics is equivalent to the one described above in the SMC section. We first perform equilibration runs for $N_{t,\text{eq}}=2 \times 10^9$ steps, and afterwards sample the equilibrium ensemble for $N_{t,\text{prod}}= 10^{10}$ steps, by extracting $10^4$ configurations equally distributed over time. This procedure amounts to around $9 \times 180$ CPU hours for $N=77$ (nine cores for roughly one week). We have performed 2 (for $N=44$, 55), 24 (for $N=66$), and 78 (for $N=77$) independent PT runs.
	
\begin{figure*}
		\includegraphics[scale=0.63]{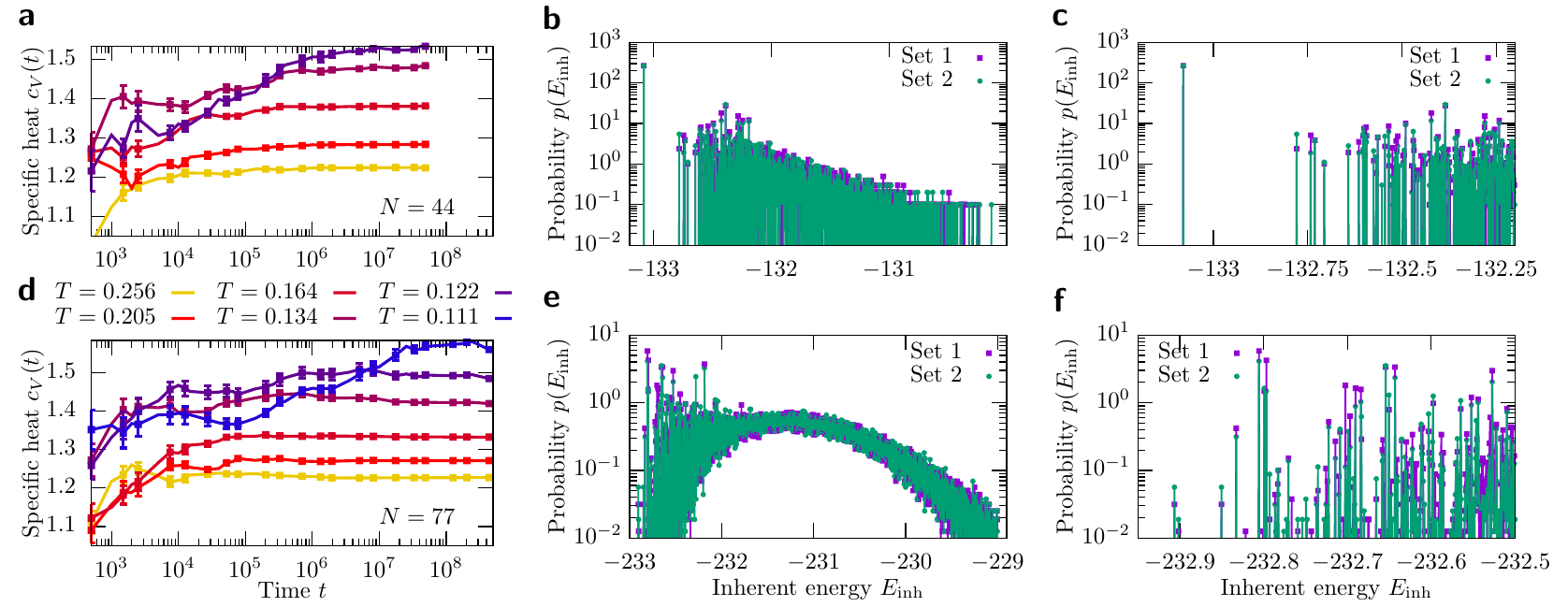}
		\caption{\textbf{Sampling configurations using parallel tempering.} \textbf{a.} Time-dependent $c_V(t)$ for different temperatures $T$ for $N=44.$ All curves reach a clear plateau, thus indicating equilibration. 
			\textbf{b.} Probability distribution of inherent energies for two independent sets for $N=44.$ Both sets yield basically the same probabilities. 
			\textbf{c.} Same as \textbf{b} but zoomed to emphasize the lowest inherent energies.
			\textbf{d/e/f.} Same as \textbf{a/b/c} for $N=77.$}
		\label{fig:S_sampling_pt}
	\end{figure*}

	In Fig.~\ref{fig:S_sampling_pt} we show an analysis of the equilibrium ensembles created by this sampling procedure. For any of the sampled temperatures, the time-dependent specific heat attains a well-defined plateau, clearly indicating equilibration (see Fig.~\ref{fig:S_sampling_pt}{a,b}). We additionally observe that two completely independent sets of configurations produce the same ground state and low energy excitations, similar to the observation in Ref.~\cite{parmar2023depleting}. We additionally establish the more demanding criterion that the states are sampled with the same probability (see Fig.~\ref{fig:S_sampling_pt}{c-f}). These ergodic sets will be the starting point of the population annealing procedure.
	
	\subsection{Population annealing}
	
	Population annealing is based on using an initial set of configurations, representing an equilibrium ensemble at temperature $T_0$ and perform successive reweighting and population resampling to create ensembles at lower temperatures~\cite{hukushima2003population,tokdar2010importance,machta2010population,gessert2023resampling}.
	
	We use the $N_c$ configurations sampled using the PT simulations described above as input for PA in an attempt to reach even lower temperatures. In population annealing step $n$, we choose a temperature $T_n < T_{n-1}$ slightly smaller than $T_{n-1}$ and assign the Boltzmann weight $W_m = \exp(-(\beta_n-\beta_{n-1} )E_{\text{pot},m})$ to each configuration $m$ in the population. Here, $E_{\text{pot},m}$ is the total potential energy of configuration $m$ and $\beta_n = (k_B T_n)^{-1}$ the inverse temperature with $k_B=1$. Afterwards, we create $\tau_m = N_c W_m / \sum_{s=1}^{N_c} W_s$ copies of configuration $m.$ To maintain an ensemble of constant size $N_c$, we use the ``systematic resampling'' scheme of Ref.~\cite{gessert2023resampling}. We then use these annealed configurations and perform short thermalization runs of $5 \times 10^3$ MD steps at temperature $T_n$, using the SMC dynamics. 
	
	The PA procedure starts at $T_0=0.122$ ($T_0=0.111$ for $N=77$) and is applied down to $T_{n_{\rm max}} = 0.001$ where $n_{\rm max}=25$. In other words, we perform 25 PA annealing steps to reach the lowest temperature investigated in this work. Temperature steps are slightly non-uniformly distributed with larger steps in the beginning, but the results were insensitive to this choice, as long as a sufficient number of small steps was used.
	
	In Fig.~\ref{fig:kauzmann}d of the main manuscript we have shown that the time-dependent specific heat $c_V(t=N_c \tau)$ attains a plateau, indicating equilibration. Here, $\tau=5 \times 10^3$ corresponds to the relaxation time between the output of two separate configurations in the PT run, which were then used as input for the PA procedure.  In Supplemental Fig.~\ref{fig:S_PA}a we observe the same for the time-dependent potential energy $\langle E_{\rm pot}\rangle_{N_c}$. We found that the specific heat is a superior measure for equilibration compared to the potential energy since its equilibration time is much longer, consistent with the findings in Ref.~\cite{jung2024normalizing}. 
	
	For $T=0.057$ we perform a reweighting (RW) procedure, which corresponds to a single population annealing step, starting from $T_0=0.111$ quenching directly to $T_1 = 0.057.$ We observe that using RW only, the equilibration time is significantly increased, and a plateau is not reached when studying $\langle E_{\rm pot}\rangle_{N_c}$ within our simulation time window. The reason for the improved performance of PA compared to RW can be inferred from the histogram of \iss, as shown in Supplemental Fig.~\ref{fig:S_PA}b,c. PA systematically leads to an increased probability of finding configurations with low inherent energies. As described above, this is ensured by the stepwise selection of configurations with low thermal energy and resampling their thermal energies. In contrast, when solely performing a single reweighting step, thermal fluctuations are blurring the selection of the proper low-temperature states, and reweighting is unable to correctly sample low temperatures. The above analysis highlights the importance of using a stepwise PA procedure to obtain the results presented in this manuscript. 
	
	More generally, we have only managed to create equilibrium ensembles at $T=0.001$ when using the combination of the three described Monte Carlo techniques. All three of them play a key role in this success, and each one of them is unable to achieve equilibration by itself. In Ref.~\cite{jung2024normalizing} we have intensively benchmarked enhanced sampling techniques. While PA alone performs well, we still found that it fell out of equilibrium at around $T_{\rm PA}=0.19,$ and thus at temperature much higher than the glass transition temperature $T_g$. Similarly, without the usage of PT, we were not able to equilibrate the system with $N=77$. As a conclusion, combining SMC/PT/PA in the order described above is vital to ensure equilibration down to $T \to 0$.
	
	\section{Configurational entropy}
	
	\label{sec_met:sconf}
	
	We have used three different techniques to evaluate the configurational entropy.
	
	\subsection{Expansion near \iss}
	
	The idea of this method is to evaluate the vibrational entropy $s_{\rm glass}$ by first performing a harmonic approximation around each \is. Afterwards, an anharmonic correction is applied to improve the results. Combined with the measurement of the total entropy, this procedure then allows to calculate the configurational entropy,
	\begin{equation}
	s_c = s_{\rm tot} - s_{\rm glass}.
	\end{equation}
	Our methodology mainly follows the lines of Ref.~\cite{berthier2019confentropy}, but we had to derive several finite size corrections which are relevant due to the small system sizes investigated in the present work, compared to earlier efforts.
	
	We start by calculating the total entropy $s_{\rm tot}$. The total Helmholtz free energy can be obtained by thermodynamic integration starting from infinite temperature $\beta =0$:
	\begin{equation}
	\beta F_{\rm tot} =  \beta F_{\rm id} +  \int_0^{\beta} \mathrm{d} \beta' \langle E_{\rm pot}(\beta') \rangle,
	\end{equation}
	with the ideal gas free energy,
	\begin{align}
	\beta F_{\rm id} = -\ln Z_{\rm id} - N s_{\rm mix},
	\end{align}
	derived from the partition sum,
	\begin{equation}
	Z_{\rm id} = \frac{1}{N! }\left( \frac{L^2m}{2 \pi \beta \hbar^2} \right)^{dN/2},
	\end{equation}
	the mixing entropy,
	\begin{equation}
	N s_{\rm mix} = \ln \left( \frac{N!}{\prod_\alpha N_\alpha!  }\right) ,
	\label{eq:s_mix}
	\end{equation}
	and the average potential energy $\langle E_{\rm pot}(\beta) \rangle.$ Here, $N$ is the total number of particles, $N_\alpha$ denotes the number of particles of type $\alpha,$ $d=2$ is the space dimension, and $L=L_x=L_y$ the linear box length. Using the relation $s_{\rm tot} = \frac{d}{2} + N^{-1} \beta \langle E_{\rm pot}(\beta) \rangle - N^{-1}\beta F_{\rm tot}$ we find the entropy per particle,
	\begin{equation}
	s_{\rm tot} = \frac{d}{2} + \frac{\ln Z_{\rm id}+\beta \langle E_{\rm pot}(\beta) \rangle}{N} + s_{\rm mix}  - \int_0^{\beta} \mathrm{d} \beta' \langle E_{\rm pot}(\beta') \rangle.
	\end{equation}
	Note that using the standard terms for $s_{\rm mix}$ based on Stirling's approximation leads to a difference $\Delta s_{\rm tot}(N=44) \approx 0.15$  compared to the above expressions. Therefore, using the full formula in Eq.~(\ref{eq:s_mix}) is essential, in particular for systems with $N \lesssim 100$.
	
	In practice, we separate the integral,
	\begin{eqnarray}
	\int_0^{\beta} \mathrm{d} \beta' \langle E_{\rm pot}\rangle &=&  \int_0^{\beta_0} \mathrm{d} \beta' \langle E_{\rm pot} \rangle + \int_{\beta_0}^{\beta} \mathrm{d} \beta' \langle E_{\rm pot} \rangle \nonumber \\ &=& I_{\rm F} + I_{\rm N}.    
	\end{eqnarray}
	For $I_{\rm N}$, we use standard numerical integration schemes, for example Simpson integration, from $\beta_0$ to $\beta$.
	For $I_{\rm F}$, we fit the temperature-dependence of the potential energy based on the high-temperature expansion~\cite{coluzzi2000lennard}:
	\begin{equation}
	\beta \langle E_{\rm pot}(\beta) \rangle  = A \beta^x + B(\beta^x)^2
	+ C (\beta^x)^3 + \cdots ,
	\label{eq:high_T_expansion}
	\end{equation}
	where $A$, $B$, and $C$ are coefficients, and $x=d/n$ with $n$ being the exponent of the repulsive power-law potential ($x=1/6$ in our case). By using this expansion, we can calculate $I_{\rm F}$ via
	\begin{equation}
	I_{\rm F} = \frac{1}{x} A \beta_0^x + \frac{1}{2x}B(\beta_0^x)^2 + \frac{1}{3x} C(\beta_0^x)^3 + \cdots .
	\label{eq:I_F}
	\end{equation}
	
	The different contributions to the total entropy are shown in Supplemental Fig.~\ref{fig:S_sconf}a. We observe that even for $N=77$ the finite-size corrections by not using Stirling's approximation account to $\Delta s_{\rm tot}(N=77)>0.03$ which is a very relevant contribution, especially at low temperatures where we seek to study the vanishing of the configurational entropy. Supplemental Fig.~\ref{fig:S_sconf}{a} also highlights that the non-trivial temperature dependence of $s_{\rm tot}$ emerges from the energetic contributions.
	
	Having calculated the total entropy, we now evaluate the glass entropy, $s_{\rm glass}$, which denotes the entropy contribution from thermal vibrations. The harmonic approximation is based on expanding the potential energy around its \is, $E_{\rm pot} \approx E_{\rm inh} + \frac{1}{2}\Delta \bm{R}^T H \Delta \bm{R}$, where $\Delta \bm{R}=\bm{R}-\bm{R}_{\rm inh}$ and $H$ is the Hessian matrix. This enables the calculation of the partition sum,
	
	\begin{eqnarray}
	Z_{\rm harm} &=& \Lambda^{-Nd} e^{-\beta { E_{\rm inh}}} \int d \bm{R}  e^{-\frac{1}2 \beta \Delta \bm{R}^T H \Delta \bm{R} } ,
	\end{eqnarray}
	where $\Lambda=\sqrt{2 \pi \hbar^2 \beta}$ is the thermal de Broglie wavelength. We do not have the mixing term because each \is exists $\prod_\alpha N_\alpha! $ times. $H$ is diagonalized by an orthogonal matrix $U$. We then obtain the eigenvalues of $H$ (sorted by ascending order), $0 = \lambda_1 = \lambda_2 = \cdots = \lambda_{d} < \lambda_{d+1} \leq \lambda_{d+2} \leq \cdots \leq \lambda_{Nd}$.
	The first $d$ eigenvalues are zero due to translational invariance.
	Using a variable transformation, $\bm{X}=U\Delta \bm{R}$, one gets
	\begin{eqnarray}
	Z_{\rm harm} &=& \Lambda^{-Nd} e^{-\beta { E_{\rm inh}}} \prod_{s=1}^{Nd} \int X_s e^{-\frac{\beta}{2}\lambda_s X_s^2} \nonumber \\
	&=& \Lambda^{-Nd} e^{-\beta { E_{\rm inh}}} \left( \prod_{s=1}^{d} \int X_s \right) \prod_{s=d+1}^{Nd} \int X_s e^{-\frac{\beta}{2}\lambda_s X_s^2} \nonumber \\
	&=& \Lambda^{-Nd} e^{-\beta { E_{\rm inh}}} N^{\frac{d}{2}} L^d \prod_{s=d+1}^{Nd} \sqrt{\frac{2 \pi}{\beta \lambda_s}} .
	\end{eqnarray}
	In the last equation we used $X_s = \sqrt{N} r^{\rm CM}_s$, where $r^{\rm CM}_s$ is the $s$-th component of the center of mass, due to the orthonormality of the eigenvectors associated with the zero modes. This gives rise to $\int d X_s = \sqrt{N}L$.
	
	Finally, we calculate the glass entropy,  $s^{\rm harm}_{\rm glass} = \frac{d}{2} + N^{-1}\beta \big( \langle E_{\rm pot} \rangle -  \overline{F_{\rm harm}} \big)$ by subtracting the free energy $\beta F_{\rm harm}= - \ln Z_{\rm harm}$ from the kinetic and potential energies. Here, $\overline{(\cdots)}$ denotes the average over independent inherent states. Consistent with previous work, we assume within the harmonic approximation $\langle E_{\rm pot} \rangle \approx \overline{E_{\rm inh}} + \frac{(N-1)d}{2}T$. Collecting all the terms, we find
	\begin{equation}
	s^{\rm harm}_{\rm glass} = \frac{1}{N} \left[ \sum _{s=d+1}^{Nd} \left( 1 - \overline{\ln (\beta \hbar \omega_s)} \right) + \frac{d}{2} \left( 1 - \ln \left( \frac{2 \pi \hbar^2 \beta}{N L^2} \right) \right)  \right]
	\label{eq:sc_harm_glass} ,
	\end{equation}
	where $\omega_s=\sqrt{\lambda_s}$.
	This term is identical to the typical expression (as used for instance in 
	Ref.~\cite{berthier2019confentropy}), including finite-size corrections of order $\mathcal{O}(N^{-1}\ln N)$.
	
	The above calculations were performed within the harmonic approximation and we have neglected anharmonic contributions, $ \langle E_{\rm anh}\rangle = \langle E_{\rm pot} \rangle -  \left( \overline{E_{\rm inh}} + \frac{(N-1)d}{2} T \right) $. To improve upon the harmonic approximation, we fit $ \langle E_{\rm anh} \rangle$ using polynomials of the form $ \langle E_{\rm anh}(T)\rangle /N = \sum_{k=2} a_k T^k.$ This allows us to evaluate the anharmonic contribution,
	\begin{equation}
	s^{\rm anh} = N^{-1} \int_{0}^T dT^\prime \frac{1}{T^\prime} \frac{ \partial \langle E_{\rm anh}(T^\prime)\rangle}{\partial T^\prime},
	\end{equation}
	which yields, $s^{\rm anh} = \sum_{k=2} \frac{k}{k-1} a_k T^{k-1}. $ 
	Finally, we collect all terms to obtain the configurational entropy:
	\begin{equation}
	s_c^{\rm harm} = s_{\rm tot} - s^{\rm harm}_{\rm glass} - s^{\rm anh}. 
	\label{eq:sum}
	\end{equation}
	
	The different contributions to $s_c^{\rm harm}$ are shown in Supplemental Fig.~\ref{fig:S_sconf}b. Importantly, we observe that the vanishing of the configurational entropy at $T=0$ is far from trivial as it results from a cancellation of three large terms in Eq.~(\ref{eq:sum}). In particular, $s_{\rm tot}$ is obtained from a thermodynamic integration starting at $T = \infty$, while $s^{\rm harm}_{\rm glass} + s^{\rm anh}$ is based on the harmonic approximation around the $T \to 0$ \is. Both contributions thus emerge from opposite and independent limits. The observation that $s_c^{\rm harm}(T \to 0) \to 0$ is thus a strong indication that we have correctly sampled the equilibrium configurations at low temperatures, and that the proposed estimate of the configurational entropy is physically meaningful. 
	
	\subsection{Expansion near equilibrium configurations: Frenkel-Ladd method}
	
	A major disadvantages of the harmonic approach described above is that it relies on the concept of \iss. While well-established, it is desirable to have an independent measure of configurational entropy. This can be achieved by applying a method close to the one proposed by Frenkel-Ladd (FL)~\cite{frenkel1984new} to compute the free energy of solids. While being physically similar to the harmonic approach, the FL method calculates vibrational contributions via a thermodynamic integration from an Einstein solid without the need to determine \iss. In this approach, the configurational entropy is
	\begin{equation}
	s_c^{\rm FL} = s_{\rm tot} - s^{\rm FL}_{\rm glass}, 
	\end{equation}
	where $s_{\rm tot}$ is the total entropy derived above. The FL glass entropy is determined from the Helmholtz free energy via thermodynamic integration from a system in which each particle in a reference configuration is harmonically trapped. The interaction energy becomes $\beta U_\alpha(\bm{R}, \bm{R}_0) = \beta E_{\rm pot}(\bm{R}) + \alpha \sum_{i=1}^N (\bm{r}_i - \bm{r}_{0, i})^2$. Here,  $\bm{R}_0=(\bm{r}_{0,1}, \bm{r}_{0,2}, ...,\bm{r}_{0,N})$ denotes the positions of the particles in a given reference equilibrium configuration. The corresponding free energy and partition function are given by 
	\begin{eqnarray}
	\beta F_\alpha &=& - \ln Z_\alpha , \\
	Z_\alpha &=& \Lambda^{-Nd} \int \bm{R} e^{-\beta U_\alpha(\bm{R}, \bm{R}_0)} .
	\end{eqnarray}
	
	Special care is required when computing the free energy for small system sizes using the Frenkel–Ladd method. The thermodynamic integration requires accurate evaluation of the mean-squared displacement from the reference configuration under a harmonic potential, as explained below. The Monte Carlo estimation of this quantity must be performed with the center of mass (CM) motion removed. Otherwise, the integration diverges when the spring constants become small. However, our goal is to estimate the glass entropy including the degrees of freedom associated with the center of mass, since this entropy must be compared with the total entropy, which also accounts for the CM contribution. For sufficiently large systems, the distinction between CM-fixed and CM-free treatments is negligible. In our case, however, where the system size is small, this distinction has a non-negligible impact on the precise determination of the configurational entropy. Therefore, it is essential to first relate the numerically accessible free energy of the CM-fixed system to that of the CM-free system, which is what we wish to obtain~\cite{frenkel2000finiteN}.
	
	In practice, we rewrite $\beta F_\alpha$ as 
	\begin{eqnarray}
	\beta F_\alpha = - \ln Z_\alpha = - \ln Z_\alpha^{\rm CM} + \ln \frac{Z_\alpha^{\rm CM}}{Z_\alpha} ,
	\label{eq:relation_CM_fixed_free}
	\end{eqnarray}
	where 
	\begin{equation}
	Z_\alpha^{\rm CM} = \Lambda^{-Nd} \int \bm{R} e^{-\beta U_\alpha(\bm{R}, \bm{R}_0)} \delta (\bm{r}^{\rm CM}) .
	\label{eq:Z_CM}
	\end{equation}
	Here $\delta (\bm{r}^{\rm CM})$ constrains the position of the center of mass, $\bm{r}^{\rm CM}=N^{-1}\sum_{i=1}^N \bm{r}_i$, at the origin $\bm{0}$ which is set by the center of mass in the reference configuration: $N^{-1}\sum_{i=1}^N \bm{r}_{0,i}=\bm{0}$. The last term in Eq.~(\ref{eq:relation_CM_fixed_free}) corresponds to the probability to find $\bm{r}^{\rm CM}$ at the origin in the CM-free ensemble, which is nothing but $L^{-d}$:
	\begin{equation}
	\frac{Z_\alpha^{\rm CM}}{Z_\alpha}  = \langle \delta(\bm{r}^{\rm CM}) \rangle_\alpha = \frac{1}{L^d} .
	\end{equation}
	This term, along with the delta function in Eq.~(\ref{eq:Z_CM}), represents a finite-size correction to the standard approach.
	
	When $\alpha_{\rm max}$ is sufficiently large, the system corresponds to an Einstein solid and we have
	\begin{eqnarray}
	\beta F^{\rm FL}_{\rm glass} &=& \beta \overline{ E_{\rm pot}(\bm{R}_0)} -\ln Z_{\rm Ein}  - \lim_{\alpha_{\rm min} \rightarrow 0} {N}\int_{\alpha_{\rm min}}^{\alpha_{\rm max}} d \alpha \Delta(\alpha) \nonumber \\ 
	&\quad& + \ln \langle \delta(\bm{r}_{\rm CM}) \rangle ,
	\end{eqnarray}
	with the partition sum,
	\begin{equation}
	Z_{\rm Ein} = \Lambda^{-Nd}  \int d \bm{R}  e^{-\alpha_{\rm max} \sum_{i=1}^N (\bm{r}_i - \bm{r}_{0,i})^2}  \delta(\bm{r}_{\rm CM}),
	\end{equation}
	and the average displacement from the reference configuration, $\Delta(\alpha)= N^{-1} \overline{\langle \sum_{i=1}^N (\bm{r}_i - \bm{r}_{0,i})^2 \rangle_\alpha^{\rm CM}}$, where $\overline{(\cdots)}$ denotes averaging over reference configurations. We fix the CM motion during the Monte-Carlo evaluation. $Z_{\rm Ein}$ can be solved analytically using the inverse Fourier transform of the delta function.
	\begin{equation}
	Z_{\rm Ein} = \Lambda^{-Nd} \int \frac{d \bm{k}}{(2 \pi)^d} \prod_{i=1}^N \int d \Delta \bm{r}_i  e^{-\alpha_{\rm max} \Delta \bm{r}_i^2 - \frac{{\rm i}}{N} \bm{k}^T \Delta \bm{r}_i } ,
	\end{equation}
	where $\Delta \bm{r}_i=\bm{r}_i - \bm{r}_{0,i}$.
	Subsequently, we perform the Gaussian integrals to get
	\begin{equation}
	Z_{\rm Ein} = \Lambda^{-Nd} \left(\frac{\pi}{\alpha_{\rm max}} \right)^{Nd/2} \left(\frac{\alpha_{\rm max} N }{\pi } \right)^{d/2} .
	\end{equation}
	This term is identical to the finite $N$-correction derived in Ref.~\cite{frenkel2000finiteN}. We finally get the glass entropy:
	\begin{eqnarray}
	s^{\rm FL}_{\rm glass} &=& \frac{d}{2} - \frac{d}{2} \left( 1 - {\frac{1}{N}} \right)  \ln \left(\frac{\alpha_{\rm max}}{\pi} \right) - \ln \Lambda^d  + \frac{d}{2} \frac{\ln N}{N}   \nonumber\\
	&\quad& + \frac{\ln L^d}{N} + \lim_{\alpha_{\rm min} \rightarrow 0}\int_{\alpha_{\rm min}}^{\alpha_{\rm max}} d\alpha \Delta(\alpha) .    
	\label{eq:s_glass_FL}
	\end{eqnarray}
	
	We evaluate $\Delta(\alpha)$ using Monte Carlo simulations and integrate the results to obtain the final term in Eq.~(\ref{eq:s_glass_FL}).  
	We also take the size dispersity into account following the methodology derived in Ref.~\cite{ozawa2018configurational}.  In practice, we compute $\Delta(\alpha)$ using swap Monte Carlo and estimate the additional mixing entropy contribution with Monte Carlo sampling that employs only particle-swap moves with no translational displacements. In our system, the influence of polydispersity is negligible, in particular at lower temperatures, so it does not affect much our results.
	
	In Supplemental Fig.~\ref{fig:S_sconf} we compare the glass entropy calculated from the Frenkel-Ladd method, $s^{\rm FL}_{\rm glass},$ the harmonic approximation evaluated above and find a very good agreement. This shows that \iss are not required to define a configurational entropy.
	
	\subsection{Enumeration of \iss}
	
	At very low temperatures, the equilibrium ensemble of thermalized configurations populates a small number of \iss. This enables us to directly enumerate them~\cite{parmar2023depleting}. Since we have direct access to the probability $p_m(T)$ to observe \is $m$ at temperature $T$, we can estimate the configurational entropy via the Shannon entropy of the probability distribution in the canonical ensemble,
	\begin{equation}
	s_c^{\rm inh}(T) = - \frac{1}{N} \sum_m p_m(T) \ln p_m(T).
	\end{equation}
	At higher temperatures we expect this measure to deviate from the true configurational entropy, since we will not be able to enumerate all \iss anymore. Therefore, we have not shown data for $s_c^{\rm inh}(T)$ above $T=0.075$ in Fig.~\ref{fig:s_c}a.
	
	\section{Overlap and Franz-Parisi potential}
	\label{met_sec:FP}
	
	The overlap $Q$ between two independently sampled configurations $\bm{R}_1$ and $\bm{R}_2$ with $N$ particles is calculated as~\cite{coluzzi1998approach}, 
	\begin{equation}
	Q(\bm{\mathcal{D}}) = \frac{1}{N} \sum_{i=1}^N F\big( \mathcal{D}_{i} \big)
	\end{equation}
	with the minimal distance between configurations $\bm{\mathcal{D}}=(\mathcal{D}_{1},\mathcal{D}_{2},...,\mathcal{D}_{N}).$ Here, $\bm{\mathcal{D}}$ is the distance vector between two configurations, $\bm{\mathcal{D}}= (|\bm{r}_{1,1} - \bm{r}_{2,1}|, |\bm{r}_{1,2} - \bm{r}_{2,2}|,...,|\bm{r}_{1,N} - \bm{r}_{2,N}|),$ where $\bm{r}_{m,i}$ is the position of particle $i$ in configuration $m.$  The function $F(x) \in [0,1]$ should be unity for very small distances and quickly decays to zero for larger distances.  We maximize the overlap $Q$ over possible translations $\mathcal{T}$, rotations $\mathcal{R}$, and mirror transformations $\mathcal{M}$ for the configuration $\bm{R}_2$. Importantly, labeling of the particles in configuration $\bm{R}_2$ is defined such that it minimizes the total distance, $\bm{\mathcal{D}}^T \bm{\mathcal{D}}$, via the Hungarian algorithm \cite{kuhn1955hungarian}, which gives rise to $\bm{\mathcal{D}}_{\rm min}$.
	We therefore find
	\begin{equation}
	\max_{\mathcal{T},\mathcal{R}, \mathcal{M}} Q(\bm{\mathcal{D}}_{\rm min}).
	\label{eq:overlap}
	\end{equation}
	This expression defines the quantity analyzed throughout this manuscript. We have chosen the specific form,
	\begin{equation}
	F(x) = \frac{1}{2} \left(  e^{ -\left(\frac{x}{a_0(T)} \right)^2 } +   e^{ -\left(\frac{x}{a_1}\right)^2 } \right) ,
	\end{equation}
	where $a_1=0.1$ and $a_0(T)$ corresponds to the square root of the plateau height of the mean-squared displacement at temperature $T.$ None of the results shown in this manuscript depend qualitatively on this specific choice of $F(x),$ but both terms together ensure a stable position of the potential minima observed in Fig.~\ref{fig:Franz_Parisi}a for the various temperatures by accounting for decreasing thermal fluctuations as $T \to 0$.  Without the first term, at very low temperatures, all overlaps would be $Q=1,$ and without the second term the small-$Q$ peak shifted strongly towards $Q=0$. 
	
	We maximize the overlap by iterating over various transformations. In two dimensions, we need to use three rotations ($\pi/2$, $\pi$ and $3 \pi /2$) and four mirror symmetries ($x=0$, $y=0$, $x=y$ and $x=-y$). We find the optimal translation by making steps of $\Delta r = 0.5\sigma_{11}.$  This implies that we need to verify $(2L/\sigma_{11})^2$ copies of configuration $\bm{R}_2$. 
	For each of these copies we apply the Hungarian algorithm to find the minimal distance vector $\bm{\mathcal{D}}$ between the copy and configuration $\bm{R}_1$. Subsequently, we subtract the average distance, i.e., ${\mathcal{D}}_i \rightarrow {\mathcal{D}}_i - N^{-1}\sum_j {\mathcal{D}}_j.$ This step explains in hindsight why it is sufficient to validate translations in coarse steps $\Delta r$: if two configurations strongly overlap, all particles will have the identical distance, which will then be removed in the subsequent correction step. Using the corrected distance vector, we calculate $Q$ and finally choose the maximum after iterating over all copies.
	
	The optimization procedure is illustrated in SI Fig.~S5a. It shows how two seemingly very different configurations are actually the same after mirroring the $x-$coordinates and shifting. The above algorithm iteratively validates various possibilities and determines the optimal transformations and shift.
	
	We extract the overlap $Q_{mn}$ by selecting $N_s=10^3$ independent configurations $\bm{R}_m$ and, similarly, $N_s$ independent configurations $\bm{R}_n$. It has been ensured that $\bm{R}_m$ and $\bm{R}_n$ originate from two independent SMC+PT+PA runs. We can thus exclude the possibility that large overlaps stem from insufficient sampling or slow decorrelation within a single run. 
	
	Finally, we calculate the quenched Franz-Parisi potential,
	\begin{equation}
	\beta V(Q) = -\frac{1}{N N_s}\sum_{m=1}^{N_s}  \ln \left( \sum_{n=1}^{N_s} \delta(Q-Q_{m n}) \right),
	\end{equation}
	where the $\delta-$function is discretized with a bin size $\Delta Q=0.01$. The term inside the logarithm is the histogram of overlap values sampled for a given reference configuration $m$, and the first sum averages over independent $m$ configurations.  
	
	In addition to the Franz-Parisi potential shown for $N=55$ in Fig.~\ref{fig:Franz_Parisi}a, we show the ones extracted for $N=44$ and $N=77$ in SI Fig.S5b,c, respectively. While all systems show a clear transition at a critical temperature $T_K$ (which depends on $N$) the potentials look quantitatively slightly different. This is mainly caused by the local minima observed in SI Fig.~S5b,c. These local minima correspond to configurations coming from related families with overlaps $Q$ which are between $Q_{\rm liq}$ and $Q_{\rm glass}$ (see Fig.~\ref{fig:States} and discussion). For $N=55$ only one family contributed at lower temperatures, giving the Franz-Parisi potential a simpler shape without local minima. 
	
	Transitions such as shown in Fig.~\ref{fig:Franz_Parisi} have only been observed for structural glass-forming liquids using biased Monte Carlo simulations~\cite{guiselin2022statistical} or systems with pinned particles~\cite{kob2013probing}. Here, we find it using a large set of independent equilibrium structures in the bulk.
	
	\section{Constructing families of low-energy states}
	\label{sec_met:families}
	
	Using Eq.~(\ref{eq:overlap}) we calculate the overlap between all the low-energy states analyzed in Fig.~\ref{fig:States} of the main manuscript. The ground state defines the first family (yellow sphere in Fig.~\ref{fig:States}a). Subsequently, we iterate over all low-energy states, sorted by energy and starting from the ground state. If the state has an overlap $ Q  < 0.8$ with all previously identified family members, a new family is created and the current state is identified as the family minimum (purple spheres in Fig.~\ref{fig:States}a). If it has an overlap $Q>0.8$ with one of the family minima it is instead identified as an excited state and included into an existing family, chosen from the state with which it has the largest overlap. 
	
	Each state is characterized by its inherent energy difference, $\Delta E_{\rm inh}= E_{\rm inh} - E_{\rm gs},$ where $E_{\rm gs}$ is the ground state energy (shown in Fig.~\ref{fig:States}b). Each family has an excitation energy $\Delta E_{\rm fam}=E_{\rm fam} - E_{\rm gs} $, where $E_{\rm fam}$ is the energy of the family minimum (shown in Fig.~\ref{fig:States}c). Finally, each state also has an excitation energy $\Delta E_{\rm intra}=E_{\rm inh} - E_{\rm fam} $, denoting its energy relative to the family minimum (shown in Fig.~\ref{fig:States}d). As a consequence, the energy difference to the ground state is formed by adding the two individual excitation energies, $\Delta E_{\rm inh} =  \Delta E_{\rm fam} + \Delta E_{\rm intra}$.
	
	To create the tree-based and hierarchical representation in Figs.~\ref{fig:States}a,b we use the k-Medoids clustering algorithm provided by the python library FasterPAM~\cite{Schubert2022}. This clustering algorithm is usable for non-Euclidean metrics such as the overlap $Q.$ Different from typical clustering algorithms, the non-Euclidean nature of $Q$ requires the most central point within the cluster to be an actual object, also called the medoid.  The total number of clusters (and thus medoids) is set to equal the number of families found using the procedure described above. In the k-Medoids algorithm the medoids are iteratively adapted to minimize the total distance of each object to the nearest medoid. We then order the states within one family according to their inherent energies. Subsequently, families themselves are clustered if the states belonging to each of the families have an average overlap $Q>0.25.$ The final order of states is thus iteratively determined by starting with the state with lowest inherent energy (i.e., the ground state in the first iteration), followed by the states belonging to the same family. Then we include all those states that belong to families which are similar to the selected family. This procedure is iterated until all states have been selected.
	
	\section{The special case of 66 particles}
	\label{met_sec:N66}
	
	In Fig.~\ref{fig:Franz_Parisi}b it can be seen that the average overlap $\langle Q \rangle$ for $N=66$ behaves differently from that of the other system sizes. This can be explained by investigating the distribution of inherent energies and the corresponding ground state and excited states (see Supplemental Fig.~\ref{fig:S_N66}). For this system size, we observe that at $T=0.035$ a majority of the configurations have an inherent energy $-199.95<E_{\rm inh}<-199.90$, and these states have a high mutual overlap as well as a significant overlap with the ground state ($ Q=0.7$). As a  consequence, the average overlap grows $\langle Q \rangle \approx 0.5$ . However, the final transition towards $\langle Q \rangle > 0.7$ when all configurations correspond to the ground state only happens at lower temperatures, $T<0.02,$ thus explaining the peculiar two-step increase in  Fig.~\ref{fig:Franz_Parisi}b for $N=66$.
	
	This behavior can also be rationalized by analyzing the configurations themselves (see Supplemental Fig.~\ref{fig:Franz_Parisi}a,c). The ground state features a pronounced symmetry along the $x=y$ and $x=-y$ axes. The configurations remain amorphous but they already remind of a quasicrystal. The excited state featured in Fig.~\ref{fig:Franz_Parisi}c has the same symmetries. One can thus conclude that $N=66$ features some kind of order or symmetry unlike the completely disordered structures seen for the other system sizes, and is therefore an outlier.

	\newpage
	
	\begin{figure*}
		\includegraphics[scale=0.95]{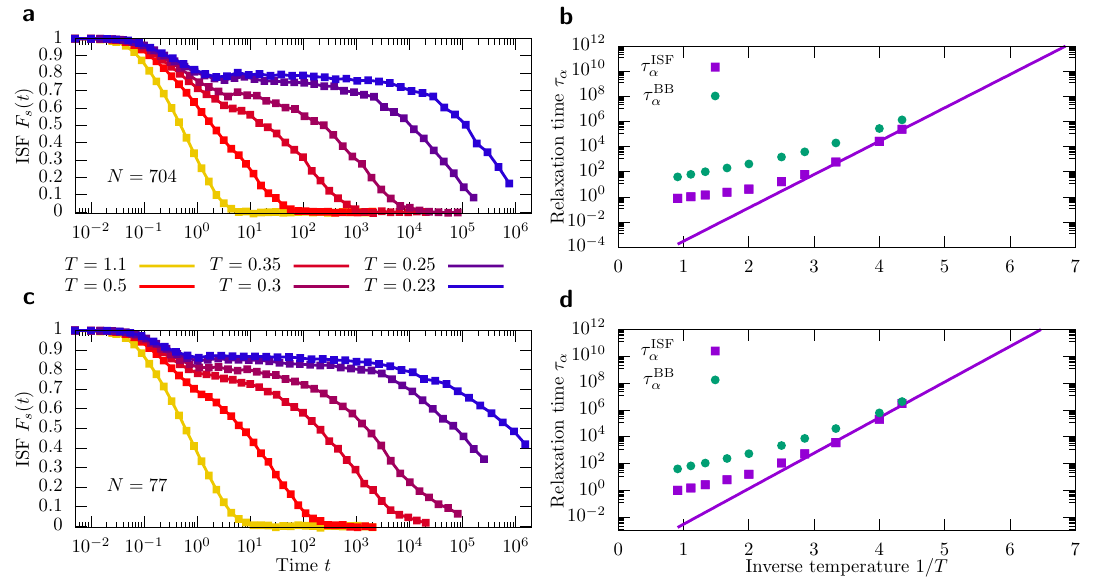}
		\caption{\textbf{Dynamics of the KA2D model.} \textbf{a.} Incoherent scattering function (ISF),  $
			F_s(t)= \left\langle N^{-1} \sum_{i} e^{-\textrm{i} \bm{q} \cdot (\bm{r}_i(t) -  \bm{r}_i(0) ) } \right\rangle,$ for $|\bm{q}|= 2 \pi \sigma_{11}^{-1}$ and $N=704.$  \textbf{b.} Temperature-dependence of the structural relaxation time $\tau_\alpha^{\rm ISF}$ defined as $F_s(\tau_\alpha^{\rm ISF})=e^{-1}.$ The purple line shows the extrapolation to $\tau_\alpha^{\rm ISF}=10^{12}$ which is a common definition of the glass transition temperature, $T_g(N=704)=0.147.$ $\tau_\alpha^{\rm BB}$ is the relaxation time of the bond-breaking correlation \cite{jung2023predicting}, showing very similar scaling behavior. \textbf{c}/\textbf{d}. Same as panels \textbf{a}/\textbf{b} for $N=77.$ The extrapolation in \textbf{d} yields,  $T_g(N=77)=0.156.$ }
		\label{fig:S_dynamics}
	\end{figure*}
	
	\newpage 
	
	\begin{figure*}
		\includegraphics[scale=0.69]{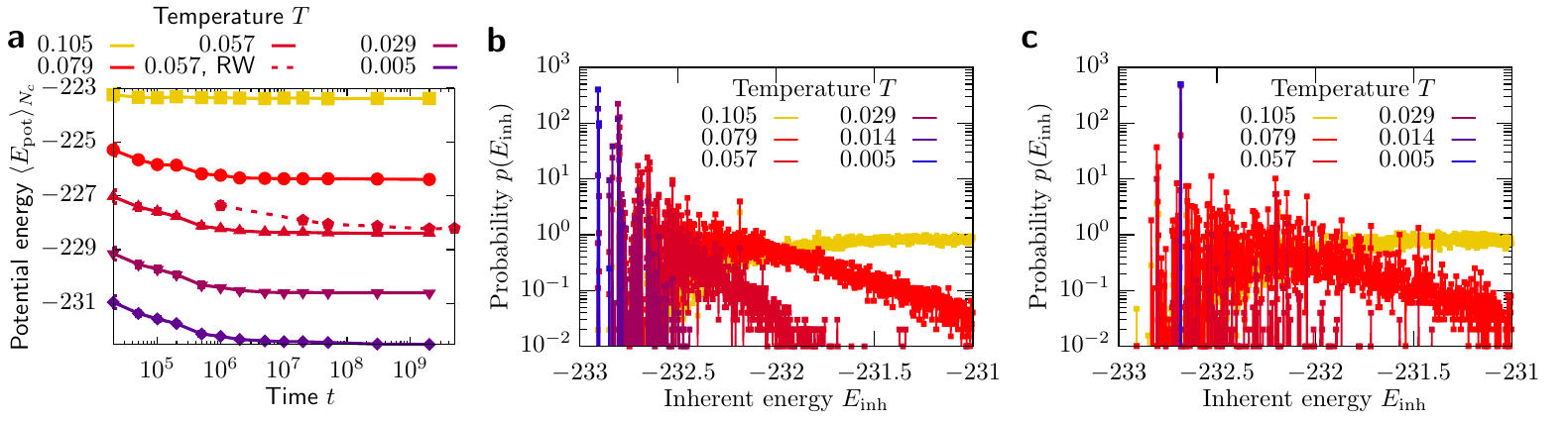}
		\caption{\textbf{Sampling configurations using population annealing (PA) and reweighting (RW) for $N=77.$} \textbf{a.} Time-dependent $\langle E_{\rm pot} \rangle_{N_c}$ for different temperatures for $N=77.$ Here, 'time' $t=N_c \tau$ corresponds to the number $N_c$ of initial configurations used for PA and RW. All curves clearly reach a plateau, thus indicating equilibration. Reweighting shows a much slower decay for $T=0.057$ (dotted line) thus indicating insufficient equilibration. \textbf{b.} Probability distribution of inherent energies for PA at different temperatures. Low energies are increasingly populated, until only the ground state is populated at $T=0.006$.  
			\textbf{c.} Same as \textbf{b} for RW. This procedure produces incorrect results at low temperatures $T < 0.07$, finally selecting a single state away from the ground state.}
		\label{fig:S_PA}
	\end{figure*}
	
	\begin{figure*}
		\includegraphics[scale=0.95]{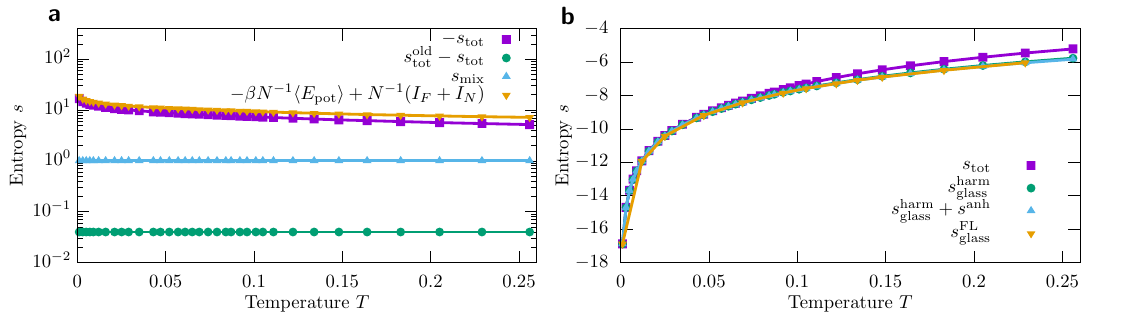}
		\caption{\textbf{Different contributions to the configurational entropy for $N=77$.} \textbf{a.} Temperature-dependence of different contributions to the total entropy $s_{\rm tot}.$ The term $s^{\rm old}_{\rm tot}-s_{\rm tot}$ describes the difference between the total entropy using finite size corrections, $s_{\rm tot}$, and Eq.~(10) in Ref.~\cite{berthier2019confentropy} without finite-size corrections, $s^{\rm old}_{\rm tot}$. All other terms are defined in Methods Section~\ref{sec_met:sconf}.  \textbf{b.} Temperature-dependence of the different contributions to the configurational entropy $s_c$.}
		\label{fig:S_sconf}
	\end{figure*}
	
	\newpage
	
	\begin{figure*}
		\includegraphics[scale=0.84]{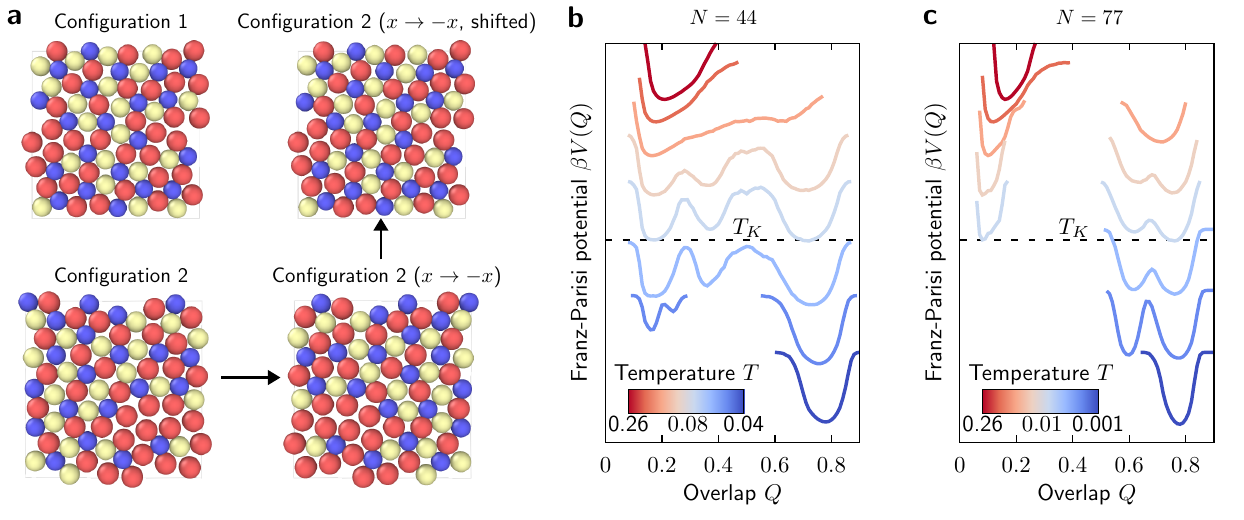}
		\caption{\textbf{Calculating the overlap $Q.$} \textbf{a.} Illustration of how the maximal overlap $Q$ is calculated. Configurations 1 and 2 are independently sampled and appear distinct. However, mirroring the $x-$coordinate and subsequently apply the shift $x\rightarrow x -2.5$, $y\rightarrow y +1.5$ reveals that the two configurations are actually very similar with overlap $Q \approx 1.$  
			\textbf{b.} Same as Fig.~\ref{fig:Franz_Parisi}a for $N=44$. 
			\textbf{c.} Same as Fig.~\ref{fig:Franz_Parisi}a for $N=77$.}
		\label{fig:S_overlap}
	\end{figure*}
	
	\begin{figure*}
		\includegraphics[scale=0.9]{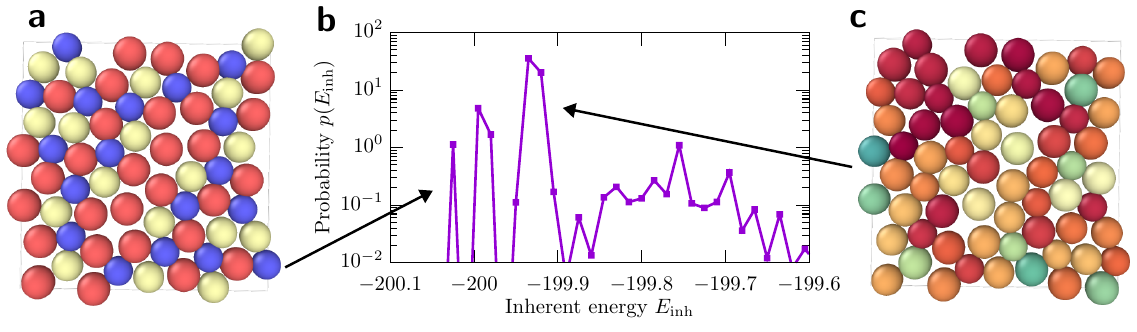}
		\caption{\textbf{The special case $N=66$.} \textbf{a.} Snapshot of the ground state for $N=66$. Color code represents the particle type. 
			\textbf{b.} Probability distribution of inherent energies for $N=66$ at $T=0.035.$ 
			\textbf{c.} Snapshot of the fourth excited state with energy $E_{\rm inh}=-199.935.$ Color code for the overlap with ground state is as in Fig.~\ref{fig:States}g.}
		\label{fig:S_N66}
	\end{figure*}
	
\end{document}